# Characterization of dust activity on Mars from MY27 to MY32 by PFS-MEX observations

Paulina Wolkenberg[1,2], Marco Giuranna[1], Davide Grassi[1], Alessandro Aronica[1], Shohei Aoki[3,1,4], Diego Scaccabarozzi[5], Bortolino Saggin[5]

[1]Istituto di Astrofisica e Planetologia Spaziali (IAPS) – Istituto Nazionale di Astrofisica (INAF), Via del Fosso del Cavaliere 100, Rome, 00133, Italy

[2]Centrum Badan Kosmicznych Polskiej Akademii Nauk, ul. Bartycka 18a, 00 – 716 Warsaw, Poland

[3]Institut d'Aéronomie Spatiale de Belgique, avenue Circulaire 3, 1180 Brussels, Belgium

[4]Department of Geophysics, Graduate school of Science, Tohoku University, Aramaki Aza Aoba 6-3, Sendai, Miyagi 980-8578, Japan

[5]Department of Mechanics, Politecnico di Milano, Campus of Lecco, Lecco, Italy

**Corresponding author**:

Paulina Wolkenberg

IAPS – INAF

Via del Fosso del Cavaliere 100

Rome, 00133, Italy

e-mail: paulina.wolkenberg@iaps.inaf.it, pwolkenberg@gmail.com

**Abstract**

We present spatial and temporal distributions of dust on Mars from Ls = 331° in MY 26 until Ls = 80° in MY33 retrieved from the measurements taken by the Planetary Fourier Spectrometer (PFS) aboard Mars Express. In agreement with previous observations, large dust opacity is observed mostly in the southern hemisphere spring/summer and particularly over regions of higher terrain and large topographic variation. We present a comparison with dust opacities obtained from Thermal Emission Spectrometer (TES) – Mars Global Surveyor (MGS) measurements. We found good consistency between observations of two instruments during overlapping interval (Ls = 331° in MY 26 until Ls = 77° in MY 27). We found a different behavior of the dust opacity with latitude in the various Martian years (inter-annual variations). A global dust storm occurred in MY 28. We observe a different spatial distribution, a later occurrence and dissipation of the dust maximum activity in MY 28 than in other Martian years. A possible precursor signal to the global dust storm in MY 28 is observed at Ls = 200° – 235° especially over west Hellas. Heavy dust loads alter atmospheric temperatures. Due to the absorption of solar radiation and emission of infrared radiation to space by dust vertically non-uniformly distributed, a strong heating of high atmospheric levels (40 – 50 km) and cooling below ~30 km are observed.

**Introduction**

Dust is one of the most variable and meteorologically important factor of the Martian atmosphere. It shows large temporal and spatial variability, and intensive radiative activity (**Heavens et al. 2011a**). For this reason, studying dust distribution and optical properties was a goal of almost every major spacecraft mission to Mars: it has been investigated by different instruments including ground-based and in-situ measurements, as well as by space-borne instruments. The dust storms on Mars, which can develop to planet-encircling and global events, are one of the most spectacular phenomena in our solar system. During a few weeks, dust is lifted into the atmosphere and starts to cover the planet almost entirely (**Strausberg et al., 2005**). The small dust storms originated as a result of strong winds, which are quite frequent on Mars, expand to larger storms and eventually likely transforming in global scale event. Globally distributed dust in the atmosphere occur in some years with a random frequency (**Strausberg et al., 2005**). During occurrences of global storms, the onset takes place in the southern hemisphere and then dust is moved to other parts of the planet supported by an intensified Hadley circulation. When dust activity starts in the northern hemisphere due to mid-latitude frontal systems, it does not disseminate globally (**Wang and Richardson, 2015; Haberle, 1986**). **Wang and Richardson (2015)** suggest that there can be also substantial southern hemisphere dust storm lifting in non-global storm years.

Dust can be lifted into the atmosphere by several mechanisms, including surface wind, dust devils and saltation, which can form local, regional and global dust storms (**Cantor et al., 2001**). These mechanisms depend on the size of the dust particles (**James, 1985**). The saltation process requires the surface wind to be around 25 – 30 m/s in order to raise the coarse particles (**Greeley et al, 1980; Cantor et al, 2001**). The saltation of dust containing large size of grains can induce the lifting of finer particles which, in turn, can activate local, regional and global dust storms, thanks to their ability to stay suspended much longer in the atmosphere than the coarse ones (**Cantor et al., 2001, Read and Lewis, 2004**). Because of a quite frequent occurrence of dust devils on the Mars surface, it was suggested that they can be responsible for lifting dust grains of all sizes (**Cantor et al., 2001**). After injection into the atmosphere, these can remain suspended for a few hours or days in the case of local dust storms and initiation phase (**Pollack et al., 1979**), and up to several weeks or months in the case of regional and decay phase of global dust storms, dependent on particle sizes (**Read and Lewis, 2004**). Dust can be carried to other locations of the planet by global circulation (e.g.,

Hadley cell, planetary waves) as well as by mesoscale and local winds (**Cantor et al., 2001**). The meridional ascending branch of Hadley cell lifts warm air up to around 40 km during southern summer and transports it to the northern hemisphere (**Cantor et al., 2001**). The southward flow of the Hadley circulation was also observed in MOC images during MY 24 from Ls = 207° to Ls = 225°, when several regional dust storms occurred (**Cantor et al., 2001**). However, the mechanism of the origin for the planet-encircling and global dust storms is still poorly understood (**Smith et al., 2002; Strausberg et al, 2005**).

Airborne dust influences atmospheric temperature and has a significant effect on the general circulation of the Martian atmosphere (**Madeleine et al., 2011**). Therefore, its radiative effect is important to be included in Global Climate Models (GCMs) (**Madeleine et al., 2011**). During day, dust absorbs solar radiation and leadings to warming of the lower atmosphere by diabatic heating. This way, dust may affect the Hadley circulation and modify thermal tides (**Zurek et al., 1992**). The increase of atmospheric dust amount produces both horizontal and vertical expansion of the meridional circulation (Hadley up- and down-welling branches) compared to a dust-free atmosphere (**Haberle et al., 1982**, Fig.11). The infrared radiation absorbed or emitted to space by dust during day and night can modulate the intensity of the radiative warming and cooling of the atmosphere (**Haberle et al., 1982; Schneider, 1983**). Dust generates more stable atmosphere. When dust loads increase, the temperatures tend to homogenize with altitude, leading to quasi isothermal profiles and thus increasing the static stability by decreasing the lapse rate (**Haberle et al., 1982; Schneider, 1983**).

In this work we investigate dust behavior during six full Martian years derived from a new dataset (**Giuranna et al., 2016; 2017, in preparation**) calculated after improvements of retrieval code described briefly in section 1. The uncertainties of atmospheric aerosol estimations during the retrieval process are discussed in section 2. A brief comparison of PFS (MEx) and TES (MGS) dust optical depths is performed for the overlapping Ls interval between MY 26 and MY 27 and presented in section 3. Moreover, we show a global map of dust distribution derived from PFS measurements carried out from MY 28 until MY32 comparable with results presented by **Montabone et al., (2017)**. The inter-annual and latitudinal variability of the total dust opacity are presented in section 4. We focus on the southern spring and summer seasons when we observe a significant dust enhancement in the atmosphere. Our studies of the global dust storm that occurred in MY 28 and of the dust activity in other Martian years are presented in section 5. Finally, as dust can induce atmospheric temperature growths or drops, we analyze this effect in terms of heating and cooling rates in section 6.

1. **Dataset and improvements of retrieval algorithm**

PFS is a Planetary Fourier Spectrometer aboard the European Mars Express mission, which carries out measurements in two spectral channels, at short-wavelength from 1.2 – 5.5 μm (8200 – 1700 cm$^{-1}$) and at long-wavelength from 5.5 – 45 μm (1700 – 250 cm$^{-1}$). The first PFS observation of the Martian atmosphere dates back to Ls = 331.18 of MY 26 (January 2004) and it is still performing measurements at the time of writing. The spectral resolution of both channels is 1.3 cm$^{-1}$ for unapodized spectra. The spatial resolutions for the short-wavelength channel and for the long-wavelength channel are 6.5 km and 11.5 km respectively, derived from the instrument field of view (FOV) which is equal to 1.52° and 2.69°, and a spacecraft pericenter altitude of 240 km. A complete description of the instrument and its radiometric performances can be found in **Formisano et al. (2005)** and **Giuranna et al., (2005a, 2005b)**.

In this work, we focus on radiation measured in nadir geometry by the long-wavelength channel, which contains information on atmospheric temperatures, surface temperatures and the column-integrated optical depth of dust and water ice. The dataset contains measurements obtained at all latitudes from pole to pole, covering all local solar times.

In order to retrieve the above quantities, we use the optimal estimation method with the Bayesian approach (**Rodgers, 2000**). The vertical axis is pressure. Relation (Eq. 1) allows us to derive the state vector $x_j$ during the iterative procedure from step $j$ to step $j+1$:

$$x_{j+1} = x_j + \left((1+\gamma)S_a^{-1} + K_j^T S_e^{-1} K_j\right)^{-1} \left[K_j^T S_e^{-1}(y - F(x_j)) - S_a^{-1}(x_j - x_a)\right] \quad [1]$$

The final state vector $x_j$ describes our best estimation of the atmospheric conditions based on the 'a priori' information taken from $x_a$, the covariance matrix ($S_a$) and the measurements $y$. In our case, the state vector $x_j$ includes the atmospheric parameters such as the atmospheric temperatures on the pressure grid, the total dust and water ice opacities, and the surface temperature to be retrieved using the dedicated retrieval code (BDM, Bounded Data Manager) described in Grassi et al. (2005). $F(x_j)$ is the forward model (radiative transfer algorithm) which calculates the synthetic spectra for the state vector $x_j$. $K_j$ are the weighting functions, namely the Jacobian of the forward model with regards to all retrieved parameters, which are calculated at each iteration step ($j$). $S_e$ is a diagonal matrix which contains the covariance of the measurement error. We use the NER (Noise Equivalent Radiance) as the uncertainty of observations. The $x_a$ 'a priori' state vector is taken from the EMCD ver. 4.2 (European Mars Climate database, Forget et al., 1999a, b) as the typical atmospheric conditions on Mars vary with local time (LT), location and season. $S_a$ – the 'a priori' covariance matrix – is derived from variances outputted by the EMCD. Recently, the retrieval algorithm got improved to build a new dataset of these parameters taking into account the dusty seasons. Namely, (1) the number of iterations is increased, (2) the 'a priori' covariance matrix is updated, (3) the Levenberg - Marquardt method is applied to the Bayesian algorithm with a value of the γ parameter varying at each iteration; γ is the stabilization parameter in the Levenberg-Marquardt method. Presently, the algorithm stops when the iterations exceed 80. Before these improvements, the variances of the 'a priori' covariance matrix ($S_a$) were taken from the EMCD model (Grassi et al., 2005). Now, we included minimum values for the standard deviations of several atmospheric and surface parameters, in case the standard deviations returned by the EMCD are too small. The standard deviations of the atmospheric temperatures are set to a minimum value of 10 K, the surface temperatures to 10 K and the aerosol total opacities to 2. If the EMCD model foresees larger variances of each matrix element, then the retrieval code assumes these value as an input. This approach allows us to better retrieve the dust opacity, especially during high opaque atmosphere conditions. The above mentioned method is used to derive the atmospheric conditions on the whole PFS database.

The information about the dust optical depth can be obtained from two spectral ranges included in the PFS thermal infrared spectra, namely 400–500 cm$^{-1}$ (25 – 20 μm) and especially 1050–1150 cm$^{-1}$ (~9.52 – 8.7 μm), where the dust shows broad absorption features. Atmospheric temperatures are derived from radiances measured in the spectral range from 550 – 800 cm$^{-1}$ (18.2 – 12.5 μm), where the strong 15-μm $CO_2$ absorption band is located. **Fig.1a** presents a fit of typical PFS spectra to synthetic ones with low and high amount of dust in the atmosphere. Atmospheric temperatures retrieved from the measurements are shown in **Fig.1b**. The two spectra have been acquired during the northern fall season (Ls = 242.7°) at around 2 pm LT and ~93° E longitude, but are separated ~38° of latitude. In this example, the atmosphere with high dust amount is always warmer than that with lower abundance of dust, up to ~15 K warmer around 20-30 km altitude. An exception is observed in the first 6 kilometers of altitude, where the dusty atmosphere profile shows lower temperatures.

2. **Uncertainty of dust (and ice) retrievals**

The total error of a retrieved parameter (a "state vector" in the Bayesian analysis) can be estimated from the total covariance matrix S: $S = \left(K^T S_e^{-1} K + S_a^{-1}\right)^{-1}$, [2] which is the sum of the covariance matrix $S_a$ for the 'a priori' value of a state vector, and the covariance matrix $S_e$ containing the statistical description of measurement errors. The weighting function matrix $K$ collects the partial derivatives of the forward model with respect to every element of state vector (additional information and discussion can be found in **Grassi et al., 2005**). The diagonal elements of this matrix are the variances of retrieved parameters, including dust opacity. By using [2] we are able to provide the standard deviations of each element for a state vector $x$, namely, temperatures for each pressure level, water ice, and dust column-integrated optical depths and surface temperatures for each spectrum analyzed during a retrieval process.

In order to present a global view of retrieval uncertainty for aerosol opacities, we selected here 29 orbits with almost 5000 PFS observations for different atmospheric conditions and aerosol loads, acquired on different locations and for different seasons and local times. **Fig. 2a, 2b,** and **2c** present the surface temperatures and the variance of the retrieved dust and water ice opacities for the selected measurements, respectively. The variance of retrieved opacities is clearly related to the values of the surface temperatures. Namely, large variances are observed for low surface temperatures (or, equivalently, for low signal-to-noise ratio spectra), as one would expect. A more in-depth analysis revealed the dataset can actually be divided in two sub-datasets, based on the surface temperatures. The temperature threshold is found to be 220 K for dust and 210 K for ice. Two different populations of standard deviations exist in the two temperature regimes. In **Figures 3 and 4** we show the histograms of the standard deviation for dust and ice retrievals. As we can see, for both aerosols, the distribution peaks around small values in the "warm regime", with a typical standard deviation ranging from ~0.02-0.06 for dust (**Fig. 3a**), and a sharp peak around 0.01 for water ice (**Fig. 4a**). On the other hand, in the "cold regime" the distribution's peak is observed at larger values, with a typical standard deviation of ~0.11 for dust (**Fig. 3b**), and 0.06 or lower for water ice (**Fig. 4b**). **Figures 3 and 4** are only intended to provide a global view of the retrieval uncertainty for aerosol opacities. This variance analysis conflates uncertainty in the retrieval with meteorological variability presented in next section (**Fig. 5 and 6**). The dataset of atmospheric parameters retrieved from PFS observations contains the standard deviation associated to each and every single retrieval.

3. **Comparison with TES, THEMIS and MCS data.**

We present a comparison of PFS retrievals with dust opacities obtained from TES measurements (**Christensen et al., 2001; Smith, 2004**). The two instruments operated simultaneously from Ls = 331° of MY 26 until around Ls = 77° of MY 27. This temporal interval gives us the possibility to make a direct comparison of the retrieved opacities. **Fig. 5** presents PFS zonal mean dust opacities as a function of TES retrievals. The PFS dataset of dust opacity is binned according to the TES grid point. The Ls, latitudinal and longitudinal bin is 5°, 3° and 7.5°, respectively. Colors represent different Ls intervals. The combined standard deviation (less than 0.1) for both instruments is plotted as a dashed line. **Fig.5** shows good consistency between dust opacities obtained from PFS and TES spectrometers. Mostly, data are distributed within the 1-σ deviation.

We also produce a global spatial map of dust distribution from MY 28 to MY 32 (**Fig.6**) to be compared with global maps obtained by **Montabone et al., (2017)**. Fig. 1 in **Montabone et al., (2017)** presents spatial dust distribution derived from collected data of TES and THEMIS spectro-imager from MY 24 to MY 26. There is a general good agreement between the two datasets. Large dust activities are found over Hellas and Argyre basins in both maps. Less dust is observed in TES and THEMIS data over the whole southern hemisphere up to around 30°N, compared to PFS data. However, the difference in the global mean values of dust opacity (~0.05) is within the uncertainty of the retrievals (**Fig.6**). Fig. 2 in **Montabone et al. (2017)** shows a global map of THEMIS and MCS dust retrievals collected from MY 28 until MY 32 (as for PFS in Figure 6). Their mean value of dust opacity (0.14) is also consistent with our results (0.16), (**Fig.6**). Moreover, regions with large dust opacities are observed over Valles Marineris and close to Isidis Planitia, which is also in agreement with our observations. Montabone et al. (2017) claim that the difference in the opacity observed over Hellas and Argyre in the two maps is due to biased statistics rather than inter-annual variability. Our results seem to be a combination of the two maps, as large dust opacity is observed either over Hellas and Argyre, and over Valles Marineris and close to Isidis Planitia (**Fig.6**), reinforcing interpretation in **Montabone et al. (2017)**.

### 4. Dust activity observed in different Martian years

**Figure 7** illustrates the seasonal variation of the zonal-mean column-integrated dust optical depth at 1075 cm$^{-1}$ collected from the end of MY 26 until the summer solstice of MY 33 (Ls = 80°). The Ls bin width is 10° and the latitude bin width is 3°. The column-integrated dust optical depth is normalized to 610 Pa, according to the formula $\tau = 610*\tau_0/P_{surf}$, where $\tau_0$ is the retrieved column-integrated dust optical depth and $P_{surf}$ is the surface pressure selected according to the location, time and season from the EMCD 4.2. In **Fig.7** gaps (lack of data) are caused by different reasons including spacecraft safe modes, spacecraft mass memory issues, solar conjunction, eclipse seasons, other spacecraft and PFS temporary issues.

The most evident feature in **Fig. 7** is the high opacity, with values larger than 2, observed in 2008 (MY 28) for almost 40° of solar longitude around the southern summer solstice (Ls = 270°), extending from the South pole up to ~50° N latitude. Dust activity, although with lower intensity (opacity not larger than ~0.5) is also observed during the other Martian years, mostly during the southern spring and summer, and mainly in the southern hemisphere and around the equator. The previous global dust event on Mars, with opacities exceeding 2, occurred in 2001 (MY 25) (**Smith et al., 2002**; **Cantor, 2007**).

We found a different behavior of the dust opacity with latitude in the various Martian years (inter-annual variations). We divided the planet in five regions according to latitude, and distinguished two regions around each polar cap, ranging from 90° to 67.5°; two at mid-latitudes from 67.5° to 31.5°; and one over the equator, ranging from 31.5°N to 31.5°S. **Figure 8** shows the variation of the dust opacity averaged over these regions as a function of solar longitude for each of the Martian years observed by PFS. The polar regions (**Figs. 8a and 8e**) show two broad peaks of dust opacity roughly centered around the two solstices.

**Montabone et al., (2015)** obtain 0.15 and 0.3 – 0.5 during summer seasons for northern and southern polar regions, respectively, which is consistent with **Fig.8 a, e**. The increase of suspended dust opacity observed in the polar regions around the local summer solstices (Ls = 90° in the northern hemisphere, and Ls = 270° in the southern one) could be related to the recession of the polar caps and the sublimation of the seasonal deposits. The sudden change of $CO_2$ state occurring at both polar regions can induce an uplifting of the dust into the atmosphere at the edges of the polar caps where strong katabatic winds (downslope winds) are also present (**Toigo et al., 2002; Doute et al., 2014**). After that, the cap continues to recede with lower sublimation rates (e.g., **Blackburn et al., 2010**) while the general circulation transports the suspended aerosols from the polar regions to lower latitudes. A minimum of opacity is then observed around the local fall equinoxes in both hemispheres for all Martian years (**Figs. 8a and 8e**). Within the winter polar vortices PFS measures ~0.2 and ~0.25 for southern and northern polar regions, respectively. This second peak of dust opacity observed by PFS in both polar regions during the local fall and winter seasons is likely due to $CO_2$ ice clouds retrieved as dust. **Heavens et al. (2011a, b, c)** came to the same conclusion to explain the presence of aerosols in the winter high latitudes observed by MCS. The distinction between atmospheric dust and $CO_2$ ice is ambiguous and difficult to achieve, because both species have overlapping features in the relevant spectral range here analyzed (7.5-25 μm). The current retrieval scheme cannot distinguish between $CO_2$ ice and suspended dust in the polar nights. Discrimination between $CO_2$ ice and dust could be obtained considering different spectral regions, but is outside of the scope of this work, and might be done in a future study. However, this also means dust cannot be completely ruled out. A possible source of dust can be the Hadley cell circulation which might allow dust to cross the vortex from above. A similar way of transport is suggested for water ice derived from the analysis of individual vertical profiles of this aerosol from MCS data (**McCleese et al., 2017**). **Montabone et al., (2015)** show medium dust opacities (~0.3) in the winter polar cap edges from TES and THEMIS dataset, especially in MY 26.

Contrary to the polar regions, a minimum of dust opacity is observed around Ls = 110° at mid and low latitudes (**Fig. 8b, c and d**). Over northern mid-latitudes (**Fig. 8b**), we also observe two maxima of dust activity repeating every year in the second half of the year, which occur somewhat earlier than the corresponding maxima observed in the southern mid-latitude regions (**Fig. 8d**). This suggests transport of dust from north to south. MY 28 is an exception, due to the occurrence of a global dust storm. Here, the maximum of dust activity is observed later in the northern than in the southern hemisphere, which suggests a probable transport of the dust by the Hadley circulation from the south to the north hemisphere (**McCleese et al., 2010; Heavens et al., 2011c**). **Wang and Richardson (2015)** described two routes for the dust, mostly oriented from north to south and east to west. The first direction (N-S) develops when the regional dust storms are observed in the northern hemisphere and the aerosols are transported zonally, concentrated into meridional channels. When the dust storms originate in the southern hemisphere, the preferred direction is from east to west, although they indicated that the dust can also move northward when occurring, e.g., over Hellas region, which is also consistent with our results (**Fig. 9**). MY 27 also shows a peculiar behavior, with large dust activity at low latitudes (averaged opacity larger than 0.2) already at Ls = 130°, which is consistent with the THEMIS observations (**Smith, 2009**).

## 5. Dust activity in dusty season

In this section, our purpose is to show dust activity in different Martian years from Ls = 180° -360°. **Fig. 8** illustrated that the latitudinal and temporal variations of dust in MY 28 are different from the other Martian years. Therefore, we separate the study of dust activity during global dust storm in MY 28 and during typical dusty conditions in other Martian years. The maps presented in **Figure 9** and **Figure 10** are derived by collecting data from all the available Martian years except for MY28 and only for MY 28, respectively.

### 5.1. Typical dust activity

In this section we aim to characterize the typical dust activity during the dusty season on Mars (southern spring and summer; **Kass at al., 2016**), with little or no interest in inter-annual variations of absolute opacities. However, the relative spatial distribution and seasonal evolution of dust is found to be very similar in the various years in the 180° - 360° range of Ls (except for MY 28), the main difference being the absolute values (see **Fig. 8**). This also ensures good spatial coverage. Maps are built on a 3° latitude- and 5° longitude-spaced regular grid. The spatial variation of the dust opacity during the southern spring and summer seasons is presented in **Fig. 9**.

We start at Ls = 180°-200° (**Fig. 9a**), when we first observe relatively large opacity (> 0.4) in some areas which is not observed earlier in the year. In this period, the dust activity mainly develops over the south-west regions of Hellas (30°S – 60°S, 45°E - 100°E), over Argyre basin (40°S – 60°S, 300°E - 330°E), and in the region between Syrtis Major and Isidis Planitia (0-30°N, 45°E -100°E). Smaller scale dust activities are also observed over a few areas in the Tharsis region (30°S - 50°N, 220°E - 280°E) and along Valles Marineris (0 - 20°S, 260°E - 320°E). These are regions of high topographic variations and large temperature variations which, in turn, generate large pressure gradients initiating the movement of air from one location to another. The larger the horizontal pressure gradient force the stronger the wind, which can efficiently lift the surface dust up to the atmosphere (**Mulholland et al., 2015; Spiga and Lewis, 2010**). Dust opacities larger than 0.3 are also observed at ~60°S latitude for almost all longitudes, resulting from the high surface wind stresses due to mainly dynamical processes including strong thermal contrast circulation ('sea breeze' on Earth) between cold polar caps and warm defrosted surface, high topographic variations (slopes) and thermal tides, and less by the sublimation of the south polar cap edges (**Toigo et al., 2002**). The total amount of atmospheric dust over these regions increases continuously during the southern spring season (**Figs. 9 b-c**). As the lifting continues, dust begins to be transported northward by the global circulation (**McCleese et al., 2010; Heavens et al., 2011c**). The dust also travels in the east-west direction, toward the north and northeast regions of Hellas in agreement with previous analyses of MOC and MARCI images (**Wang and Richardson, 2015**). Our results also show wide regional dust activities (high dust optical depth) occurring every year over specific areas, and in particular in the region between Syrtis Major and Isidis Planitia (0-30°N, 60°E-120°E) and between Xanthe Terra and Meridiani Planum (15°S-15°N, 300°E-360°E). In these regions, persistently high values of dust opacity are observed during most of the summer spring and the early summer seasons (**Figs. 9b-d**).

Significant amounts of dust are also observed closer to the South Pole and especially around the perihelion (Ls = 251°; **Figs. 9d-f**). The maximum extent of the dust activity occurs every year in the seasonal range Ls = 240° to 260°, when dust opacity is observed over the whole southern hemisphere, and up to 30°N. The largest opacities (~0.8) are also observed in this period. At the same time, areas of persistently low values of the dust opacity (< 0.1) are found over some regions between 30°N and 60°N, especially during the Ls = 240° to 300° interval (**Fig.9d–f**).

The spatial and seasonal distribution of dust presented in **Figure 9** and in **Figure 7** is in good agreement with the analysis of zonal-mean 50 Pa daytime temperature retrievals from TES/MGS and MCS/MRO performed between MY 24 and MY 32 (**Kass at al., 2016**). Similarly, to our results, the authors found large regional-scale dust storms with similar characteristics repeating every Martian year and labeled the storms as A, B, and C in seasonal order. The "A storm" is a regional-scale or planet encircling southern hemisphere dust event occurring at most southern latitudes. It tends to initiate around middle southern spring and has a moderate duration, typically in the 205°-240° Ls range, always over by the southern summer solstice. The A – type dust activities are clearly visible in PFS data presented in **Figs. 9b-c**. The "B storm" is a southern polar event which typically starts just after perihelion and reaches its peak around the southern summer solstice (**Kass et al., 2016**). The B – type dust activities are also consistently observed by PFS as shown in **Figs. 9d-f**. Our results show that the end of the A - type dust activity cannot be clearly defined as these mid-latitudes storms merge with "B storms" originating in the south polar region, especially in the 220°-260° Ls seasonal range (**Figs. 9c and 9d; Kass et al., 2016**). The "C storms" are either regional-scale or planet encircling southern hemisphere dust events, except they are very short, mostly occurring between 305° –320° Ls (**Kass et al., 2016**). Large dust opacity is observed in most of the southern hemisphere (**Fig. 9g**), but these events are generally weaker and less extended than A - type dust activities. C – type dust events can be observed in **Figures 9g-h** and are best seen in the zonal-mean dust opacity presented in **Fig.7** with the Ls bin of 10°. They are observed each year by PFS, in the same season and latitudinal range as seen by **Kass et al., (2016)**, but with different intensity. The "A" and "C" dust storms often include "flushing" dust storms, as defined by **Wang et al., (2003)**, that start in the northern hemisphere and cross the equator, where they occasionally initiate new areas of dust lifting. The dust activity reduces in late southern summer and early fall seasons (**Figs. 9h and 9i**). Moderate dust optical depths are observed regionally each year over south of Lucus Planum (30°S, 180°E) and, again, over Xanthe Terra (**Fig. 9h**).

PFS observations show a continuous growth of the dusty areas with time, from Ls 180° to 260° (**Figs. 9a-d**). The dust activity starts over specific regions in the southern hemisphere, and subsequently expands over most of the southern hemisphere, and up to 30°N where wide regional dust activities are also observed. Dust on Mars can develop in three different styles which are referred to as ''consecutive dust storms'', ''sequential activation'', and ''merging'' **(Wang and Richardson, 2015)**. During different development phases dust can manifest different combinations of styles. This can lead to overlap in time and to grow into larger scale. Even if the temporal resolution of PFS maps (20° of Ls) does not allow us to resolve dust storm types, the results in **Fig. 9** resemble those obtained by **Wang and Richardson (2015)**. Namely, at Ls = 180° - 200° (**Fig. 9a**), relatively high dust opacity is found in separate locations over Hellas, at 60°S toward west of Hellas (Noachis Terra) and close to Isidis Planitia (0 - 15°S; 60° - 80°E), Argyre basin, and east side of Valles Marineris (5°S - 5°N; 310°E - 330°E). Subsequently, wide regions of relatively high dust opacity are observed, including and extending over most of the above separate locations (**Fig. 9b-c**), suggesting a 'merging' style storm development as described by **Wang and Richardson (2015).**

Examples of 'consecutive' storm might also be observed in **Figure 9**. In **Fig. 9c** the dust from the south of Argyre is transported westward, although high dust opacity is persistently observed over that area (**Fig. 9b and c**). A similar situation takes place also over region close to the east side of Valles Marineris (Margaritifer Terra; 0°N, 300-320°E). As we can see in **Fig. 9c**, dust is transported toward Meridiani Planum (0°N, 0-20°E) and northward from Margaritifer Terra, but high abundance is still observed over the origin place. This resembles 'consecutive' feature of dust storm where the dust activates along the route and keeps the region dusty for several sols. Again, as the 'consecutive' storms last only several sols (**Wang and Richardson, 2015**), the current temporal resolution of PFS maps does not allow us to fully support this interpretation.

### 5.2. Global dust storm in MY 28

The onset of a global dust storm could be characterized by the coalescing of multiple regional storms (**Cantor, 2007**). The global storms encircle a hemisphere of the planet in a matter of days, and may also spread dust to both hemispheres within a few weeks. This spreading of the dust is commonly referred to as the "expansion" phase of the storm which also usually involves new dust lifting areas as was seen in the 2001/MY25 global storm (**Strausberg et al., 2005**). When the peak opacities reach a few optical depths, the storm enters a quasi-exponential „decay" phase (**Murphy et al., 1990**), that lasts typically over 100 Martian sols. A planet-encircling, in turn, transforming to global dust storm occurred on Mars in 2007 (MY 28). We found the seasonal and spatial evolution of dust activity in this MY has a peculiar behavior compared to a typical Mars year without a global dust storm.

In MY28 during southern spring, we observe the dust opacity over some places (more than two regions) gets larger than 0.6 (**Fig.10b**). This happens already at Ls = 200-235° (**Fig. 10b**), when significant amounts of dust are lifted up in the atmosphere over the south polar cap edge regions (65°S - 70°S), west of Tharsis Montes (15°S – 0°S; 245°E - 275°E), and west of Hellas (65°S - 50°S; 30°E - 50°E) in Noachis Terra. This event could be a precursor signal to the global dust storm. We illustrate this by plotting the probability distribution of retrieved opacities in these particular regions during Ls = 200° – 235° in MY 28 and for other MYs (**Fig.11**, grey and black lines, respectively). Histograms are normalized to the total number of measurements (Hellas – 122; Tharsis - 98 and south polar cap edge – 388 in MY 28) and are presented with the bin size of 0.1 dust opacity (**Tab.1**). The histograms relative to the typical Mars years (black curves in **Figure 11**) are clearly peaked at a value of 0.2-0.3 of dust opacity over Hellas, and of 0 – 0.1 dust opacity over the other regions. The dust optical depths observed in MY 28 are larger by 0.2 – 0.3 compared to other MYs in all considered regions (green curves in **Figure 11**). The largest variation (0.3) of peak dust opacity observed in MY 28 takes place over Hellas. The Tharsis and Southern polar cap edge regions show also some number of measurements with dust opacities larger than 0.6 in MY 28.

We perform a chi-square test between the two histograms in order to evaluate if the differences between MY28 and the other years are statistically significant, at the 0.05 level of significance (α). Our null hypothesis assumes that two distributions are similar (not statistically, significantly different). Our alternative hypothesis is that two histograms for each region are significantly different or, in other words, that the differences between MY 28 and the other years are statistical significant. For this purpose, we use a generalization of the classical chi-square test for comparing weighted and unweighted histograms presented by **Gagunashvili (2010)**, originally developed by **Fisher (1924)**. We compare the two histograms for each region by calculating the chi-square statistics for two different sample sizes (unweighted histograms) according to the formula (5) in **Gagunashvili (2010)**:

$$\chi^2_{m-1} \cong \frac{1}{N_1 \cdot N_2} \sum_{i=1}^{m} \frac{(N_2 \cdot n_{1i} - N_1 \cdot n_{2i})^2}{n_{1i} + n_{2i}} \quad [3]$$

Where:

$N_1$ – the total observed number of events for histogram 1 (other MYs)

$N_2$ – the total observed number of events for histogram 2 (MY 28)

$i$ – number of bin

$m$ – the total number of bins

$n_{1i}$ – the total observed number of events for *i*-bin in histogram 1

$n_{2i}$ – the total observed number of events for *i*-bin in histogram 2.

This formula [3] is widely applied to test the hypothesis of homogeneity and has approximately a $\chi^2_{m-1}$ distribution (**Gagunashvili, 2010**). This statistic is used when we have histograms with unweighted entries which is our specific case. Since the alternative hypothesis is that the two histograms are "different", we use the classical procedure to test our hypothesis considering the two-tailed $\chi^2_{m-1}$ distribution. The calculated $\chi^2$ is then compared with the table of $\chi^2$ for the two–tailed distribution at 0.05 level of significance (**Johnson and Kuby, 2011**). We have 10 bins in histograms, so our degrees of freedom are 9. The critical value of $\chi^2$ for *m-1* (9 degrees of freedom) is equal to 2.7 for area in left-hand tail and 19.0 for area in right-hand tail. For the considered regions (Hellas, Tharsis and south pole cap edge) we obtain $\chi^2$ = 42, 81 and 54 using [3], respectively. All of these values are greater than the critical values thus our null hypothesis can be rejected. This means that the differences between MY28 and the other years are statistically significant for all regions (**Fig.11**).

The "precursor" storm may or may not have anything to do with the subsequent global dust storm and the actual beginning of it. They are distinct in time and the "precursor" storm dissipated months before the MY28 storm truly began. Unfortunately, we only have sparse data in the 235°-270° Ls seasonal range of MY 28 (**Fig. 10c**), which prevents us to map the spatial distribution of dust in this period. **Smith (2009)** interpreted the significant increase of dust opacity observed by THEMIS at Ls = 260° as the onset of the global dust storm. Visible imagery suggests that the storm itself started around Ls = 267°. MARCI images also show a large flushing storm occurred earlier in MY 28 southern spring which appeared very much like the flushing storm that ultimately spawned the MY28 global storm (**Wang and Richardson, 2015**).

With respect to a typical Martian year, the maximum of dust activity in MY 28 occurs later in the year, between 270° and 305° of solar longitude, when high amounts of dust persist over most of the tropical and sub-tropical regions (**Figure 10d**). In this period, the total dust opacity still exceeds 2 in some locations (before data binning). Large dust opacities (up to 1.5 or more) are also observed in the Ls interval of 305° to 340°, especially over the southern tropics (**Figure 10e**). This is very different from what we observe during a non-global dust storm year, where the total dust opacity is typically lower than 0.2 (**Fig. 9h**). Consistent to THEMIS observations (**Smith, 2009**), PFS observations show the maximum activity of the MY 28 global dust storm is confined between low northern and mid southern latitudes. For this season, we also find large differences in terms of dust opacities from orbit to orbit. We found it is due to the different local times (LT) of such observations, and higher values of dust opacity are observed during daytime. The local time variation of dust opacities will be considered in the next paper.

## 6. Effect of dust on atmospheric temperatures

We used temperature vertical profiles and column-integrated dust optical depth retrieved by PFS to investigate the influence of dust on atmospheric temperatures, which is expected to be particularly evident during a global dust storm event (**Zurek, 1978**).

### 6.1. Heating and cooling rates for selected measurements during high, moderate and low amount of dust in the atmosphere

In this section we estimate the cooling and heating effects due to dust in the infrared range (~9 μm) and in the visible range (~ 0.67 μm), respectively. In our analysis we neglect the impact of $CO_2$ on atmospheric heating and cooling rates. The $CO_2$ absorbs only 1% of solar radiation, producing cooling and heating rates around 4 - 5 K/day when sun is in zenith (**Moriyama, 1974; Savijarvi et al., 2005**). The trace gases have also a negligible effect on heating and cooling rates when compared to dust.

The volumetric heating rate $Q$ (or heat power per unit volume [W/m$^3$]) depends on changes of temperature in the atmospheric slab in time (**Sanchez-Lavega, 2010**):

$$\frac{dT}{dt} = \frac{Q}{\rho \cdot c_p} \quad [4]$$

where: $\rho$ - density of air; $c_p$ – specific heat at constant pressure

The incident solar radiation is absorbed by radiatively active species producing heating of atmosphere ($Q_{solar}$), whereas the thermal emission leads to cooling of the atmosphere in the infrared region ($Q_{IR}$). The heating and cooling rates can be calculated by the following expressions (**Sanchez-Lavega, 2010**):

$$\frac{Q_{solar}(t,p)}{\rho \cdot c_p} = \pi \cdot \frac{d\tau_\lambda(p)}{dp} \cdot \left(\frac{R_{sun}}{r(t)}\right)^2 \cdot [B_\lambda(T(5780K))] \cdot \frac{g}{c_p} \cdot \exp\left(-\frac{\tau_\lambda(p)}{\cos(\theta)}\right) \quad [5]$$

$$\frac{Q_{IR}(T,p)}{\rho \cdot c_p} = -2 \cdot \pi \cdot \frac{d\tau_\lambda(p)}{dp} \cdot [B_\lambda(T_{atm}(p))] \cdot \frac{g}{c_p} \int_0^1 \exp\left(-\frac{\tau_\lambda(p)}{\cos(\theta)}\right) d\cos(\theta) \quad [6]$$

where: $\tau_\lambda(p)$ – dust optical depth at given pressure $p$ and wavelength $\lambda$; $T_{atm}(p)$ – atmospheric temperature at given pressure $p$; $R_{sun}$ – radius of the Sun; $r(t)$ – distance from the Sun to Mars at given time $t$; $B_\lambda(T)$ – Planck function at temperature $T$; $g$ – gravity acceleration on Mars; $\theta$ – solar zenith angle.

A large variety of dust amount in the atmosphere has been observed during the global dust storm in MY 28. We calculated vertical profiles of heating and cooling rates by means of the formulas above for selected PFS measurements with high, moderate and low dust loads. In particular, we selected four measurements from MEx orbits 4510, 4471, 4328 and 4428, one measurement for each orbit, with retrieved column-integrated dust optical depths of 1.73, 1.46, 0.41, and 0.16, respectively. Details on each measurement are provided in **Tab. 2**. The associated temperature profiles of these measurements are shown in **Figure 12**. We present temperature profiles with a vertical sampling grid coarser (~5 km) than that actually used in the retrieval process (~1 km). During retrieval, the atmosphere is sampled along the vertical direction with a constant step in logarithm of pressure. This sampling grid shall not be confused with actual vertical resolution of retrieval: this latter quantity accounts also for the finite width of weighting functions and varies with measurements and altitudes. Our calculations are based on the approach provided by **Rodgers (2000)** where vertical resolution or spread $w(z)$ is given by the formula:

$$w(z) = \sqrt{\frac{\int A(z,z') \cdot (z'-c(z))^2 dz'}{\int A(z,z') dz'}} \quad [7] \quad , \text{where}$$

$$c(z) = \frac{\int z' \cdot A(z,z') dz'}{\int A(z,z') dz'} \quad [8]$$

C(z) is the 'mean' altitude at nominal peak of A(z,z') - averaging kernels. We neglect negative values of A(z,z') assuming 0 values. Calculations of vertical resolution have been performed either for 6 representative measurements from orbit 362 under standard atmospheric condition with low content of aerosol, and for the temperature profiles shown in **Fig.12**. **Figure 13** presents the 'typical' vertical resolution for standard atmospheric condition (average from orbit 362), and the vertical resolution of the temperature profiles used for the calculation of heating and cooling rates, where large variation of dust content occurs. The 'typical' vertical resolution is very similar to the vertical resolution for the low dust opacity (0.16) profile of orbit 4428 (triangles and asterisks in Figure 13), and varies between 4 km within lowest 10 km of altitude, to 12 km at 80 km of altitude. As shown in **Figure 13**, the vertical resolution decreases (the higher the spread, the lower the resolution) with increasing dust content. However, even in the most dusty atmosphere, the spread is lower than 10 km for altitudes below 20 km, and increases to ~17 km at 50 km of altitude for largest dust opacities (> ~1.5).

In our calculations, we use the vertical distribution of dust opacity derived from measurements by the Mars Climate Sounder aboard the Mars Reconnaissance Orbiter (MCS – MRO) (**McCleese et al., 2007**). In particular, we make use of the approximate formula for the dust vertical distribution derived from the MCS dataset of dust opacities described in **Heavens et al. (2011a, Eq.15)**. Coefficients for this formula have been kindly provided by the authors (N. Heavens, private communication). The available MCS dataset includes coefficients for zonally-averaged profiles binned in 5° of latitude and 5° of Ls for different MYs, except for MY 28. At the time of writing, only one dust profile is available for MY 28. Being an averaged zonal-mean profile from 30°S to 30°N at Ls = 280°, it is poorly consistent with the selected PFS observations used in this analysis for MY28 (see **Tab. 2**). For this reason, we also make use of MCS zonally averaged dust profiles reconstructed using the provided coefficients for MY 29, which are largely available in the considered seasonal (240°-275° Ls) and latitudinal (25°S-45°S latitude) range. The relevant MCS profiles for MY 29 and that for MY 28 are shown in **Figure 14a**. We only considered MCS profiles with a goodness of fit $R^2 > 0.98$ (**Heavens et al., 2011a**). Among those MCS profiles, we either selected four "best" profiles (i.e., as close as possible in time and location to the four selected PFS observations), and one "typical" (e.g., most recurrent) dust profile for MY 29 (**Fig. 14a**). We use the four selected PFS temperature profiles $T_{atm}(p)$ (**Figure 12**) and the six MCS dust profiles $\tau(p)$ (four "best" and one "typical" profile from MY 29, and the averaged zonal-mean profile from MY 28) to calculate the heating and cooling rates according to formulas [5] and [6]. The MCS profiles are first normalized to the corresponding PFS column-integrated dust opacity [dimensionless]. In order to do this, we first convert the MCS dust vertical distributions (density-scaled opacity [$m^2/kg$]) into opacity [$m^{-1}$] by dividing them by an exponential density profile (the atmospheric density at the surface is also provided in the MCS dataset for each dust profile), and then calculate the normalization factors by integration over the whole atmospheric column. Examples of normalized MCS dust profiles used in our calculations are shown in **Figure 14b**. In order to calculate heating rates by means of Eq. [5], we also convert opacities from infrared to visible using the ratio of extinction efficiency factors at 0.67 μm and 9 μm $Q_{ext}(0.67\mu m)/Q_{ext}(9\mu m) = 1.84$ (**Madeleine et al., 2011**). The conversion ratio depends on dust particle size. We use the particles with $r_{eff}$ =1.65 μm and variances $v_{eff}$ = 0.35 with dust refractive indices from Wolff et al., (2009). This is consistent with the conversion ratios showed by **Madeleine et al. (2011)** in Fig.2.

We calculated the heating (H) and cooling (Q) rates using [5] and [6], and then summed up both quantities at each pressure level (net values, Q+H). The results are shown in **Figure 15**, for the "best" (**Fig. 15a**) and the "typical" (**Fig. 15b**) MCS dust profiles in MY 29, and for the mean MCS dust profile during the global dust storm of MY 28 (**Fig. 15c**), as described above. The qualitative effect of dust on the thermal structure is similar in all our calculations with different assumptions of dust profiles. We always observe a significant net heating of the atmospheric layers just above the peak altitude of dust opacity (**Figs 14 and 15**) due to absorption of the visible solar radiation, and a net cooling in the first two or three atmospheric scale heights due to radiative cooling. The net heating rate increases with total dust opacities. Even low amount of dust (opacity around 0.15) produces a net heating rate of several degrees Kelvin per day, while relatively high opacities (> ~1.5) can heat the atmospheric layers by 40 K/day or more. These results are in good agreement with theoretical calculations by **Moriyama (1975)**. When we use a fixed dust profile (e.g., the "typical" dust profile in MY 29) for all PFS observations (**Table 2**), we get similar results in all cases. Namely, the net heating is always peaked at an altitude of ~ 40km (**Fig. 15b**), which is about 15 km higher than that of the peak of the density-scaled dust opacity (**Fig. 14a**).

Atmospheric cooling by the dust is particularly evident close to the surface and in the lower atmospheric layers (**Figure 15**). The intensity of cooling rate depends on the dust opacity and cannot be neglected in the lower levels within dusty atmosphere. At a dust storm event, cooling in the infrared regions is almost as strong as heating due to absorption of the incident solar radiation by dust, and the thermal structure of the Martian atmosphere is essentially determined by dust alone. Our results are in good agreement with previous theoretical calculations. **Moriyama (1974)** showed that in lower portion of dust layers, cooling due to the IR radiation of dust becomes one of the most predominant term in heat budget. In dusty atmospheres, cooling rates of 60-80 K/days can be observed close to the surface (**Moriyama, 1974**).

The PFS profiles shown in **Figure 12** clearly show the temperature is higher in dusty atmosphere than in low-dust case, especially above 20 km. For the lower layers the temperature increase is still evident, although the slope reduces considerably, due to both radiative cooling and blocking (shadowing) of solar radiation by suspended dust. Indeed, airborne dust also reduces the down-welling solar flux effectively, producing an 'anti-greenhouse' trend (cooling at the surface, warming within the atmosphere). This is particular evident in the PFS temperature profile with the largest dust opacity, where the atmospheric temperature between 5 and 20 km is up to 20 K lower than the other cases. Airborne dust particles shadow the surface from sunshine through scattering and absorption of solar radiation; hence they always tend to cool down the surface. Indeed, surface temperatures for the four considered measurements show a gradual decrease with dust opacities (**Tab.2**). In particular, two measurements (orbits: 4471 and 4428) among the selected four have comparable solar conditions (insolation) associated with LT and latitude. The difference in surface temperatures (~ 40K, **Table 2**) between two measurements clearly shows dust effect. This means that the surface can cool down by several tens of degrees due to suspended dust at certain total opacity in the atmosphere during day. This result is consistent with **Fig. 15b, c** where the net cooling rate for the orbit 4471 is around 40 - 45 K for the atmosphere close to the surface. This validates our calculations of heating and cooling rates using dust vertical profiles derived from MCS observations. On the other hand, absorption by dust acts as a local heat-source. Also, the surface-reflected radiation is absorbed and this absorption increases rapidly with surface albedo and dust amount. In general, the net effect on the planet may therefore be either cooling or heating, depending on the optical properties of the surface, the atmosphere and the dust (**Savijarvi et al., 2005**). The anti-greenhouse trend features we observe during the Martian dust storm in MY 28 is analogous to terrestrial SW cloud radiative forcing (**Read and Lewis, 2004**).

### 6.2. Uncertainty of heating and cooling rates

As there are no simultaneous retrievals of temperature profiles by PFS and dust profiles reconstructed from MCS data, we expect the main source of uncertainty in our calculations of heating and cooling rates is the assumption of the vertical dust distribution. Although calculations with different dust profiles lead to similar qualitative interpretations (see text above), we observe differences in the net heating rates as large as 10-15 K/day when using different MCS dust profiles in MY29 (**Figures 15a and 15b**). The largest heating and cooling rates are obtained by using the averaged dust vertical distribution of MCS retrieved during the global dust storm of MY28 (**Fig. 15c**), where the maximum of dust opacity is also observed at higher altitudes (~45 km, **Fig. 14b**). Previous theoretical calculations of heating rates presented by **Zurek (1978)** and **Savijarvi et al. (2005)** showed values of 80 K/day and 70 K/day, respectively, for heavy loads of dust in the atmosphere, which are in good agreement with our results. In general, the quantitative values strongly depend on the optical properties of dust and its vertical distribution. For a given dust profile, additional sources of errors in the estimates of the heating and cooling rates are the uncertainties in the PFS retrievals of temperature profiles and dust opacities. Under the assumption that the dust vertical distribution derived from the MCS dataset represents the real atmospheric condition without errors, the influence these two sources of uncertainty can be estimated using basic propagation of errors principles. The total derivative of Q and H is then composed of partial derivatives of each variable, which has its own uncertainty. For the PFS temperature profiles we have applied the retrieval error presented in **Grassi et al. (2005)** and **Wolkenberg et al. (2009)**. The error on the PFS dust opacity is assumed to be 0.06. We used the mean value of dust standard deviation for surface temperatures > 210 K (**Fig. 3a**) because the retrieved surface temperatures for the selected cases are always larger than 250 K (**Fig. 12, Tab.1**). The final errors are shown as error bars in **Figure 15**. As we expected, they are smaller than the effects due to different dust vertical distributions (**Fig. 15**).

## 7. Summary and conclusions

In this study, we describe the spatial and temporal distributions of dust in the Martian atmosphere from Ls = 331° in MY 26 to Ls = 80° in MY 33. Our analysis of PFS observations treats separately the global dust storm occurred in MY 28 and dust storms in the other MYs. We find that regions with high topographic variations such as Hellas, Argyre, Syrtis Major, Isidis Planitia, a few areas in the Tharsis region, and along Valles Marineris are mostly locations for onset of high dust optical depths. Dust starts rising up from these regions and increases continuously during the southern spring season (**Figs. 9b-c**). In MY28, the dust activity develops over the south polar cap edge, west of Tharsis Montes (15°S – 0; 245°E - 275°E) and west of Hellas (65°S - 50°S; 30°E - 50°E) in Noachis Terra, starting from Ls = 200° with total dust contents greater than 0.6. We identify this dust event as a possible precursor of the global dust storm occurred later in MY28. As the lifting continues, dust begins to be transported northward by the global circulation up to around 30°N, and eastward, to eventually cover the whole southern hemisphere. Our results suggest that small (regional) dust storms originated in the northern hemisphere can expand southward, as also reported by **Wang and Richardson (2015)**. That the maximum of dust activity in MY 28 is observed around 20° of Ls later than in the other Martian years (Ls = 240° - 260°). Contrary to other Martian years, quite high dust abundance (dust opacity ~ 0.8) is still present in the atmosphere in late winter (Ls = 340°) of MY 28.

The different stages of dust evolution in "typical" Martian years are very similar to those observed by TES and MCS instruments (**Kass et al., 2016**). All types of storms (A, B, and C) mentioned by **Kass et al. (2016),** occurring in the same seasonal range and with almost the same duration (**Fig.7 and 9**), can be recognized in PFS data.

We investigate the influence of dust on atmospheric temperatures in terms of heating and cooling rates. By using dust vertical distributions derived from MCS data we note a strong heating of atmosphere above the dust peak, and strong cooling in the first two or three scale heights. The intensity and vertical distribution of net heating and cooling rates depend on total dust loads and its vertical profiles in the atmosphere.

**Acknowledgments**

This work has been performed under the UPWARDS project. This project has received funding from the European Union's Horizon 2020 research and innovation programme under grant agreement No633127.

Madeleine J.-B., F. Forget, E. Millour, L. Montabone and M. J. Wolff, (2011), Revisiting the radiative impact of dust on Mars using the LMD Global Climate Model, *J. Geophys. Res.*, Vol. 116, No. E11010, doi:10.1029/2011JE003855

McCleese, D. J., J. T. Schofield, F. W. Taylor, S. B. Calcutt, M. C. Foote, D. M. Kass, C. B. Leovy, D. A. Paige, P. L. Read, and R. W. Zurek, (2007). Mars Climate Sounder: An investigation of thermal and water vapor structure, dust and condensate distributions in the atmosphere, and energy balance of the polar regions, *J. Geophys. Res.*, 112, E05S06, doi:10.1029/2006JE002790

McCleese, D. J., N. G. Heavens, J. T. Schofield, W. A. Abdou, J. L. Bandfield, S. B. Calcutt, P. G. J. Irwin, D. M. Kass, A. Kleinbohl, S. R. Lewis, D. A. Paige, P. L. Read, M. I. Richardson, J. H. Shirley, F. W. Taylor, N. Teanby and R. W. Zurek (2010), Structure and dynamics of the Martian lower and middle atmosphere as observed by the Mars Climate Sounder: Seasonal variations in zonal mean temperature, dust and water ice aerosols, *J. Geophys. Res.*, 115, E12016, doi:10.1029/2010JE003677.

McCleese, D. J., A. Kleinbohl, D. M. Kass, J. T. Schofield, R. J. Wilson and S. Greybush (2017). Comparisons of observations and simulations of the Mars polar atmosphere, VI Mars Modeling and Observations workshop, Granada, Spain.

Montabone L., F. Forget, E. Millour, R. J. Wilson, S. R. Lewis, B. Cantor, D. Kass, A. Kleinboehl, M. T. Lemmon, M. D. Smith, M. J. Wolff, (2015). Eight-year climatology of dust optical depth on Mars, *Icarus*, 251, 65 – 95.

Montabone L., B. Cantor, F. Forget., D. Kass, A. Kleinboehl, M.D. Smith, M.J. Wolff, (2017). On the dustiest locations on Mars from observations, VI Mars Modeling and Observations workshop, Granada, Spain.

Moriyama S., (1974), Effects of Dust on Radiation Transfer in the Martian Atmosphere (I) – On infrared radiative cooling, *J. Meteorol. Soc. Jap.*, vol. 52, p. 457-462
27

**Figures captions**

Figure 1. (a) Examples of PFS LWC spectra with moderate and low amount of dust in the atmosphere. Solid lines represent fits to the spectra. Measurements are plotted in dashed lines. (b) Temperature profiles retrieved from the measurements presented in Fig.1a.

Figure 2. Surface temperatures (a), variance of dust opacities (b), and variances of water ice opacities (c), for 29 orbits selected in MY 28. See text for more details.

Figure 3. Histogram of standard deviations of retrieved dust opacities for (a) surface temperatures > 220 K, and (b) surface temperatures < 220 K.

Figure 4. Histogram of standard deviations of retrieved water ice opacities for (a) surface temperatures > 210 K, and (b) surface temperatures < 210 K.

Figure 5. Comparison of zonal mean dust opacities obtained from TES and PFS measurements in MY 26 and MY 27 for intervals: Ls = 330° – 340° (black), Ls = 340° – 350° (dark purple), Ls = 355° – 10° (dark blue), Ls = 10° – 15° (blue), Ls = 15° – 30° (light blue), Ls = 30° – 60° (green), Ls = 60° – 65° (light green), Ls = 65° – 75° (yellow), Ls = 75° – 80° (orange). A combined standard deviation is plotted with a dashed line.

Figure 6. A global spatial map of dust distribution from MY 28 until MY 32 obtained from PFS measurements.

Figure 7. Zonal mean of dust opacities for 6 Martian years at 1075 cm$^{-1}$. Latitude bin is 3° and the Ls bin is 10°. Red color is for dust opacities larger than 0.5. The actual maximum of zonal-mean dust opacity observed during the global dust storm of MY 28 is ~2.15.

Figure 8. Total dust opacities for different Martian years as a function of Solar Longitude (Ls) averaged for several latitude ranges: (a) 90°N – 67.5°N; (b) 67.5°N – 31.5°N; (c) 31.5°N – 31.5°S; (d) 31.5°S – 67.5°S; (e) 90°S – 67.5°S.

Figure 9. Spatial maps of total dust opacities with a topography contour: a. Ls = 180° - 200°, b. Ls = 200° - 220°, c. Ls = 220° - 240°, d. Ls = 240° - 260°, e. Ls = 260° - 280°, f. Ls = 280° - 300°, g. Ls = 300° - 320°, h. Ls = 320° - 340°, i. Ls = 340° - 360°, j. Ls = 0° - 20°. The maps have been built by averaging data from all MYs investigated in this analysis, except for MY 28.

Figure 10. Spatial maps of total dust opacities with a topography contour for the global dust storm in MY 28 (daytime observations) during: a. Ls = 165° - 200°, b. Ls = 200° - 235°, c. Ls = 235° - 270°, d. Ls = 270° - 305°, e. Ls = 305° - 340°, f. Ls = 340° - 15°.

Figure 11. Probability distribution of retrieved opacities for Hellas (65°S - 50°S; 30°E - 50°E), Tharsis (15°S – 0; 245°E - 275°E) and South polar cap edge (65°S - 70°S) during Ls = 200° – 235° in MY 28 (grey line) and for other MYs (black line). Histograms are normalized to the total number of measurements and are plotted with the bin size of 0.1 dust opacity.

Figure 12. PFS temperature profiles selected from latitude region between 25°S to 45°S and Ls interval 240°-275° in MY 28 used for calculations of heating and cooling rates.

Figure 13. A 'typical' vertical resolution for one of temperature profiles under standard atmospheric condition (orbit 362) and vertical resolutions (spread) of temperature profiles used for calculations of heating and cooling rates.

Figure 14. (a) MCS dust vertical profiles [$m^2/kg$] selected for latitude region from 25°S to 45°S during Ls interval from 240° to 275° in MY 29. They are all zonally averaged. The selected profile considered as a "typical" for the selected region and time is plotted with diamonds. The MCS dust vertical profile in MY 28 at Ls = 280° averaged for latitudes from 30°S to 30°N is plotted as a dashed line (b) MCS vertical density-scaled opacities normalized to PFS total dust opacities in Table 1. Solid lines show the "best" MCS profiles in MY29 (as close as possible in time and location to the four selected PFS observations. See text for more details). Dashed lines are for the averaged zonal-mean profile in MY28 presented in (a) (see text for more details).

Figure 15. Net heating and cooling rates calculated for (a) "best" and (b) "typical" MCS dust profiles in MY; (c) mean MCS dust profile during the global dust storm of MY 28, averaged in the region from 30°S to 30°N at Ls = 280°. See text for more details.

Table 1. Total number of measurements for specific locations in Ls = 200° - 235°.

| Regions | MY 28 | Other Martian years |
|---|---|---|
| Hellas | 122 | 182 |
| Tharsis | 98 | 535 |
| South polar cap edge | 388 | 1390 |

Table 2. Properties of selected PFS measurements in MY 28.

| Orbit | Ls | LT | Location | Dust opacity at 1075 cm$^{-1}$ | Surface temperature [K] |
|---|---|---|---|---|---|
| 4510 | 273° | 12.13 | 27°S, 117°E | 1.73±0.06 | 249.9 |
| 4471 | 266° | 12.18 | 42°S, 343°E | 1.46±0.06 | 256.8 |
| 4328 | 241° | 14.22 | 40°S, 357°E | 0.41±0.06 | 296.6 |
| 4428 | 259° | 12.82 | 34°S, 259°E | 0.16±0.06 | 301.3 |

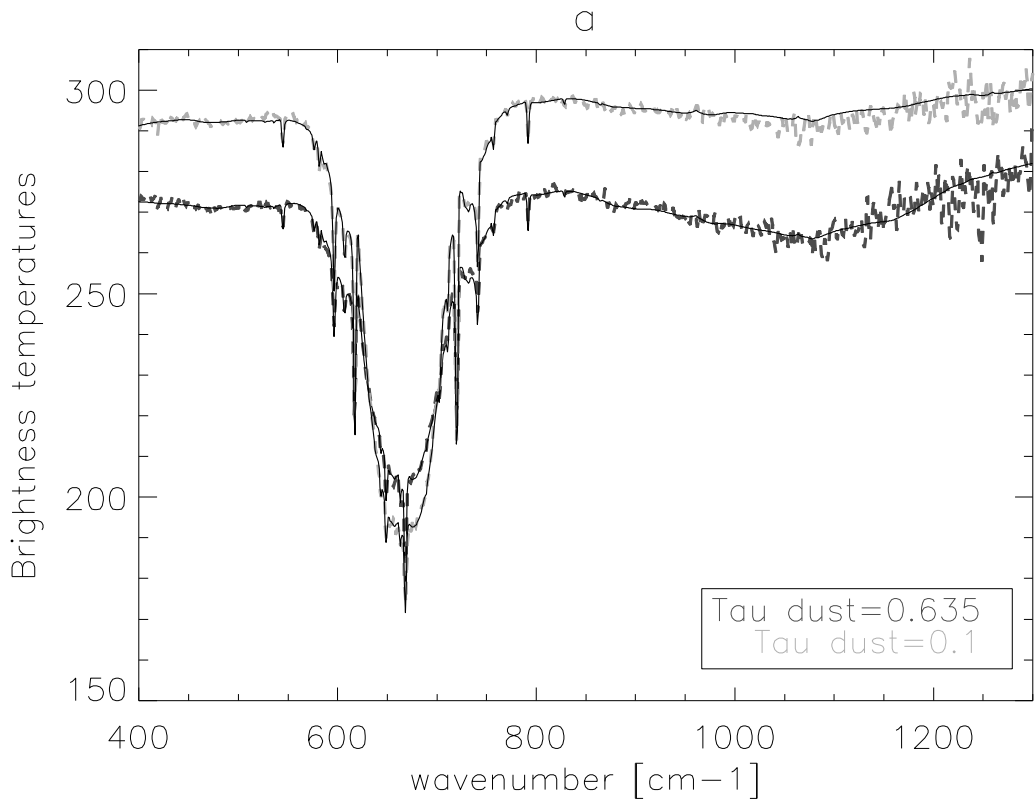

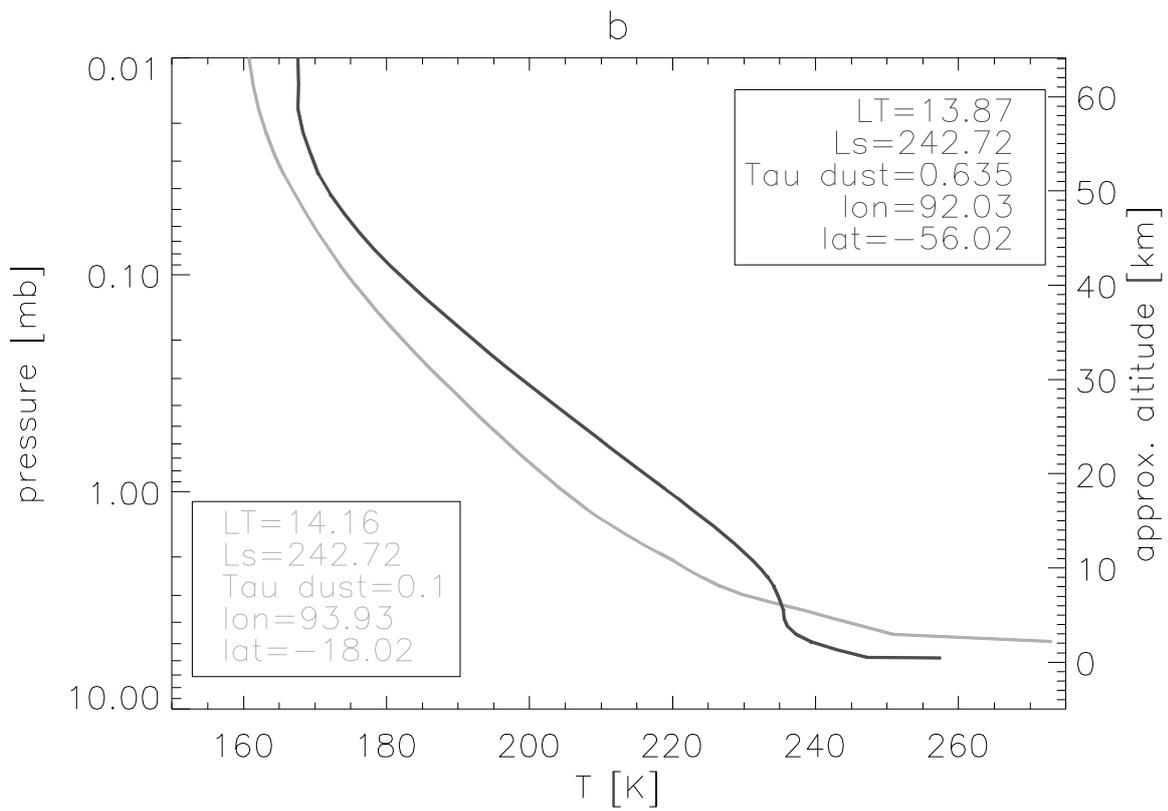

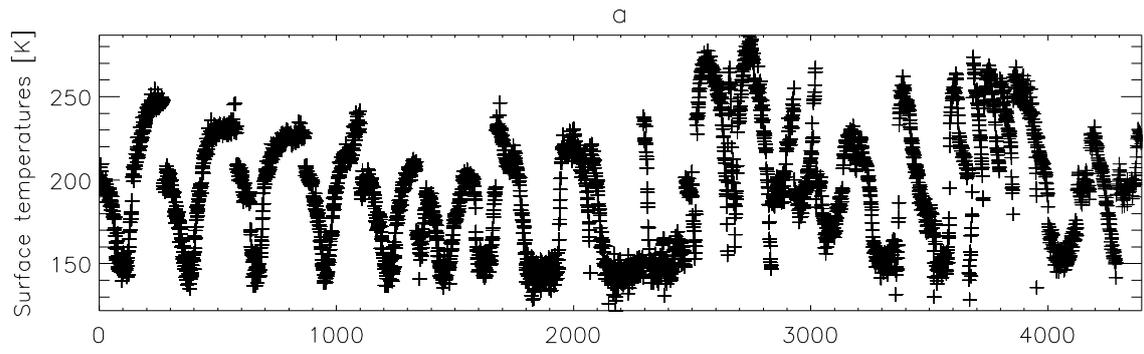
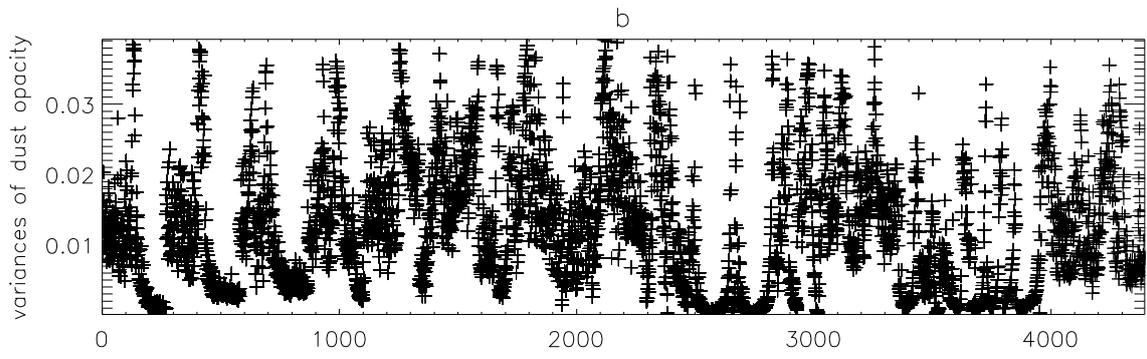
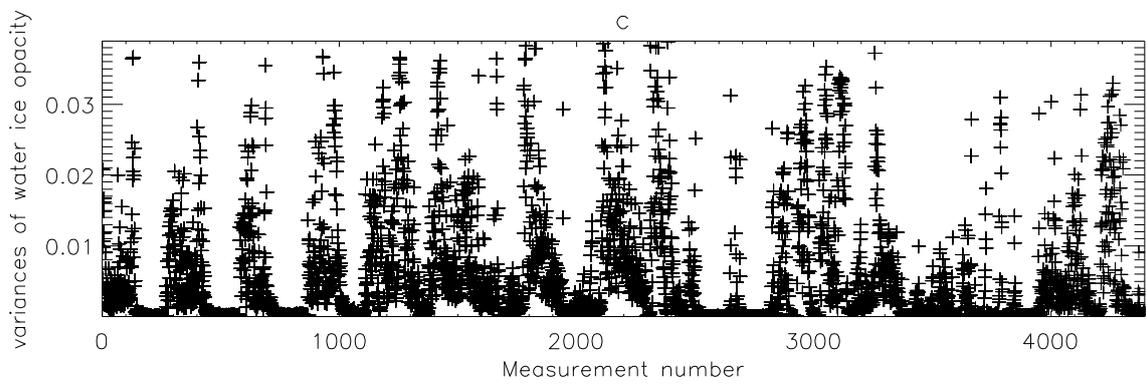

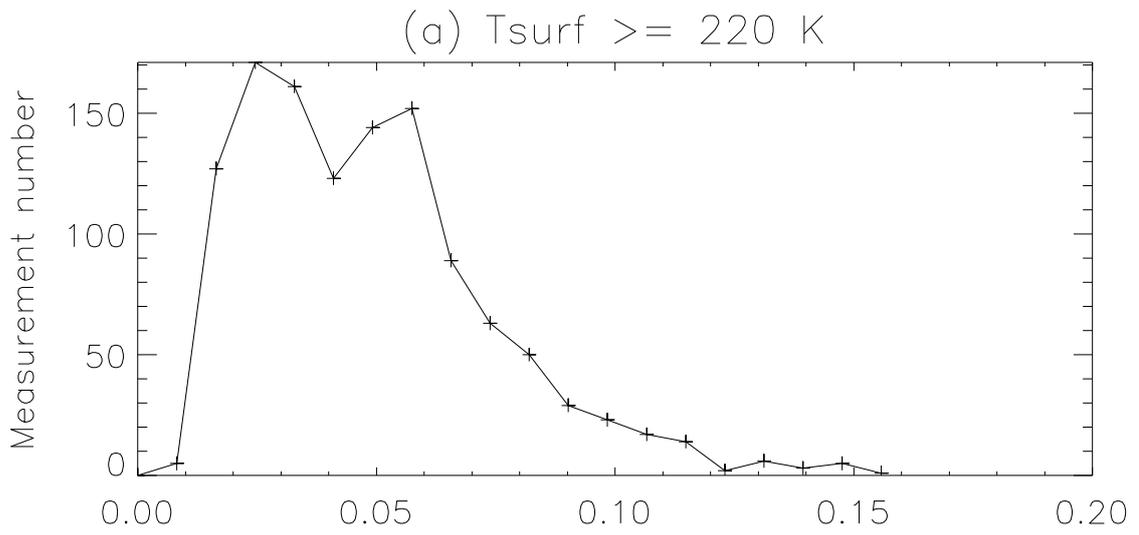

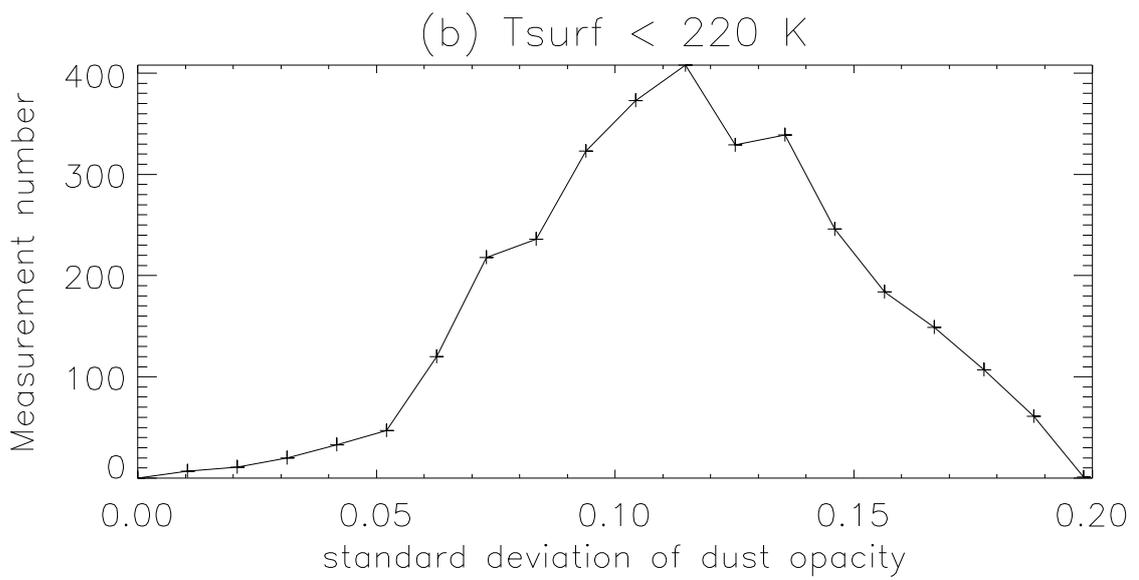

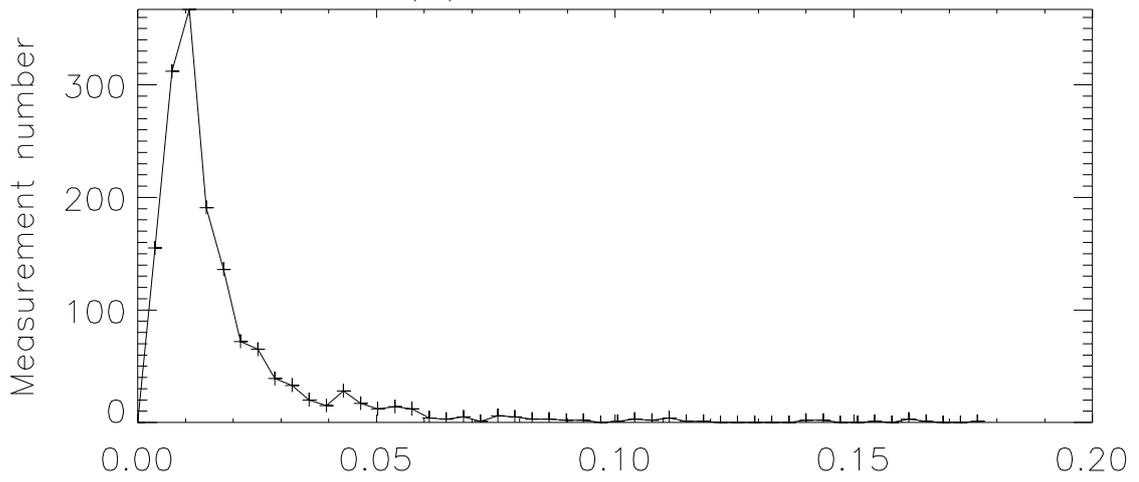

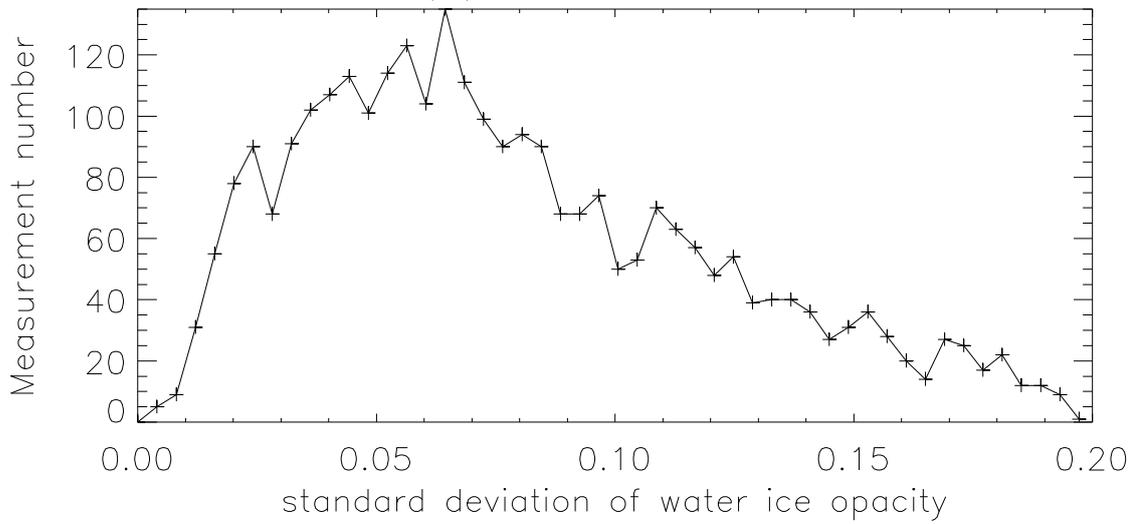

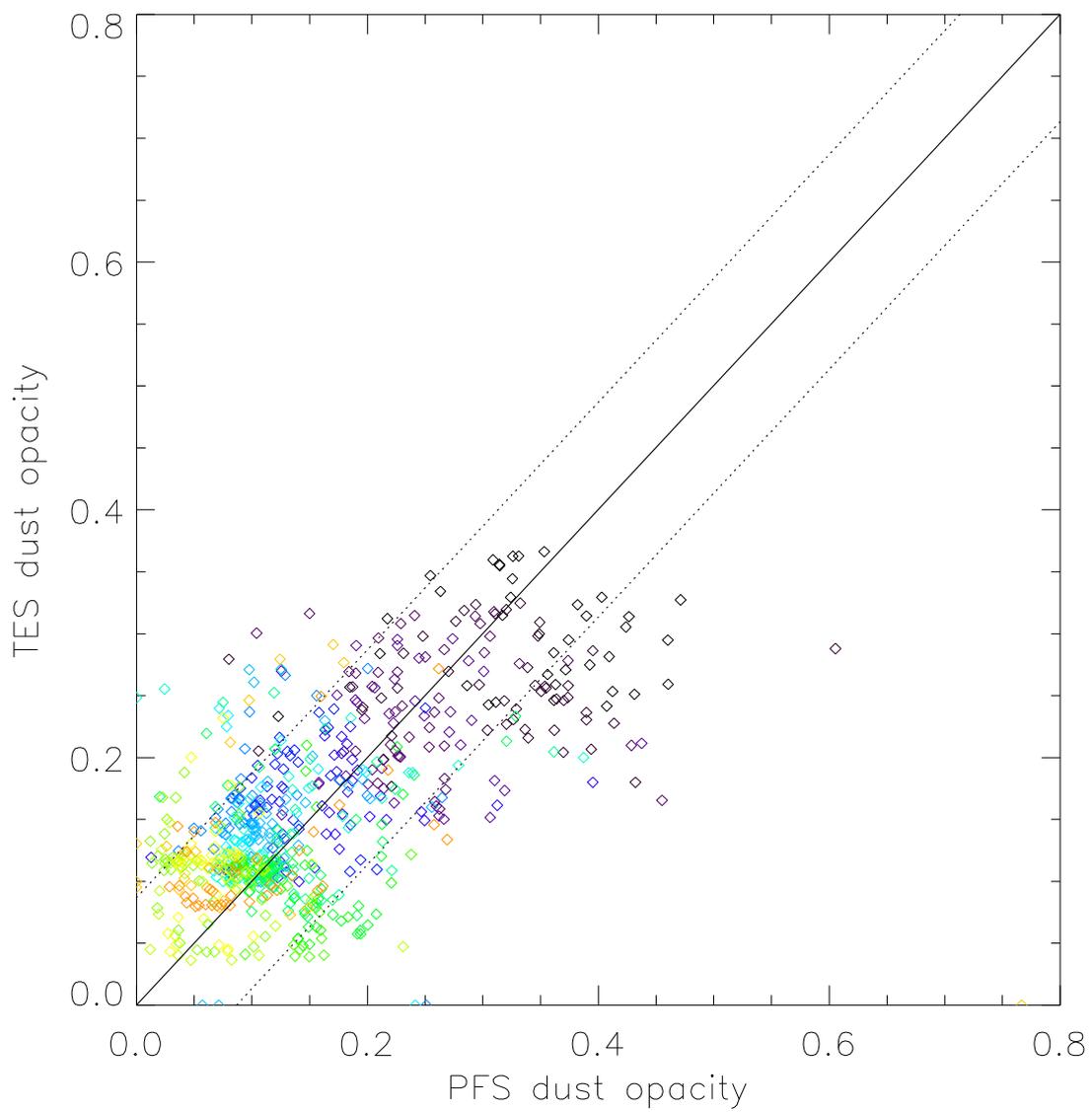

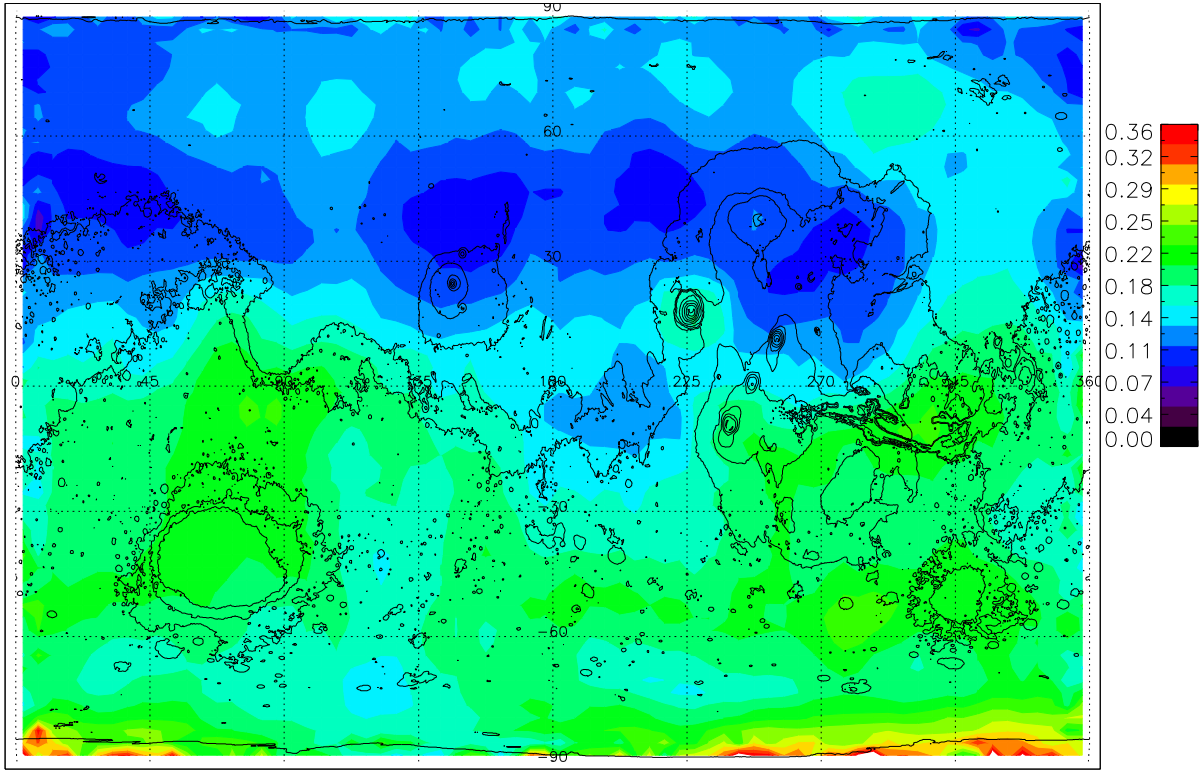

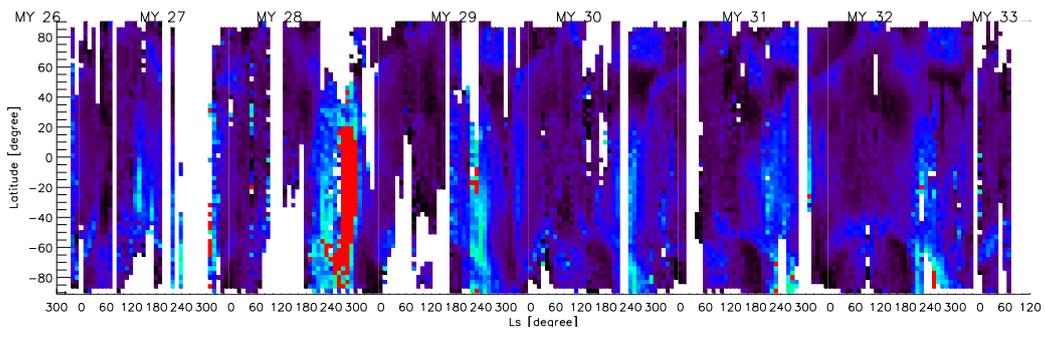

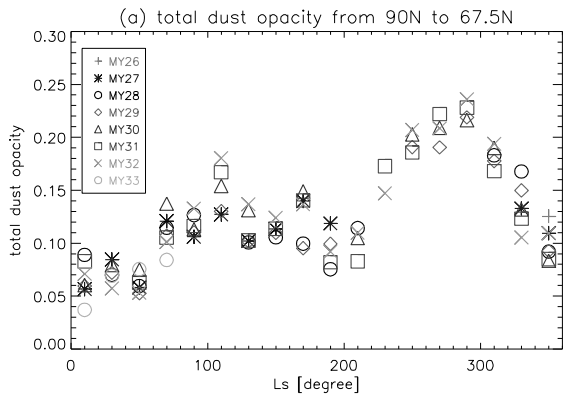
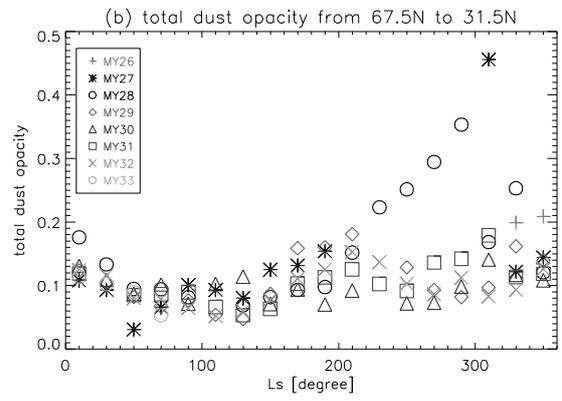
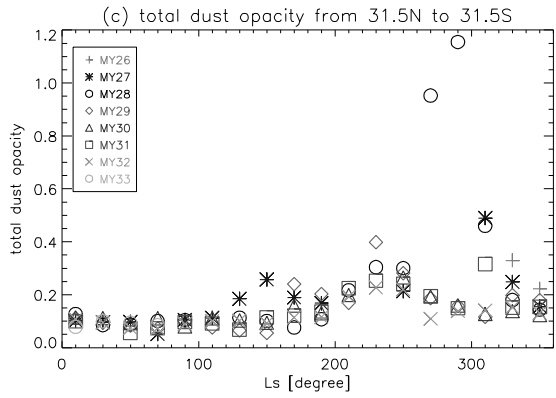
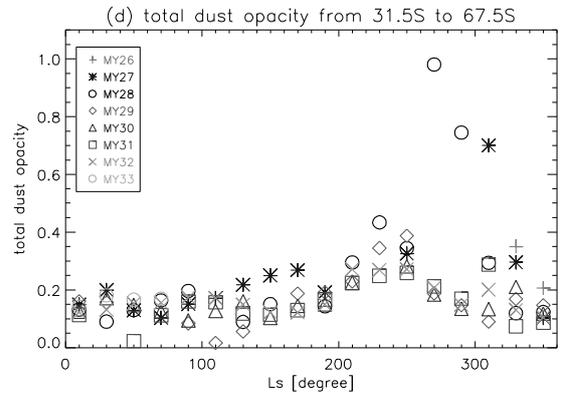
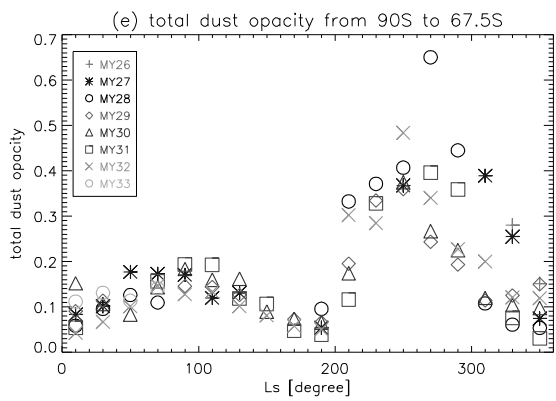

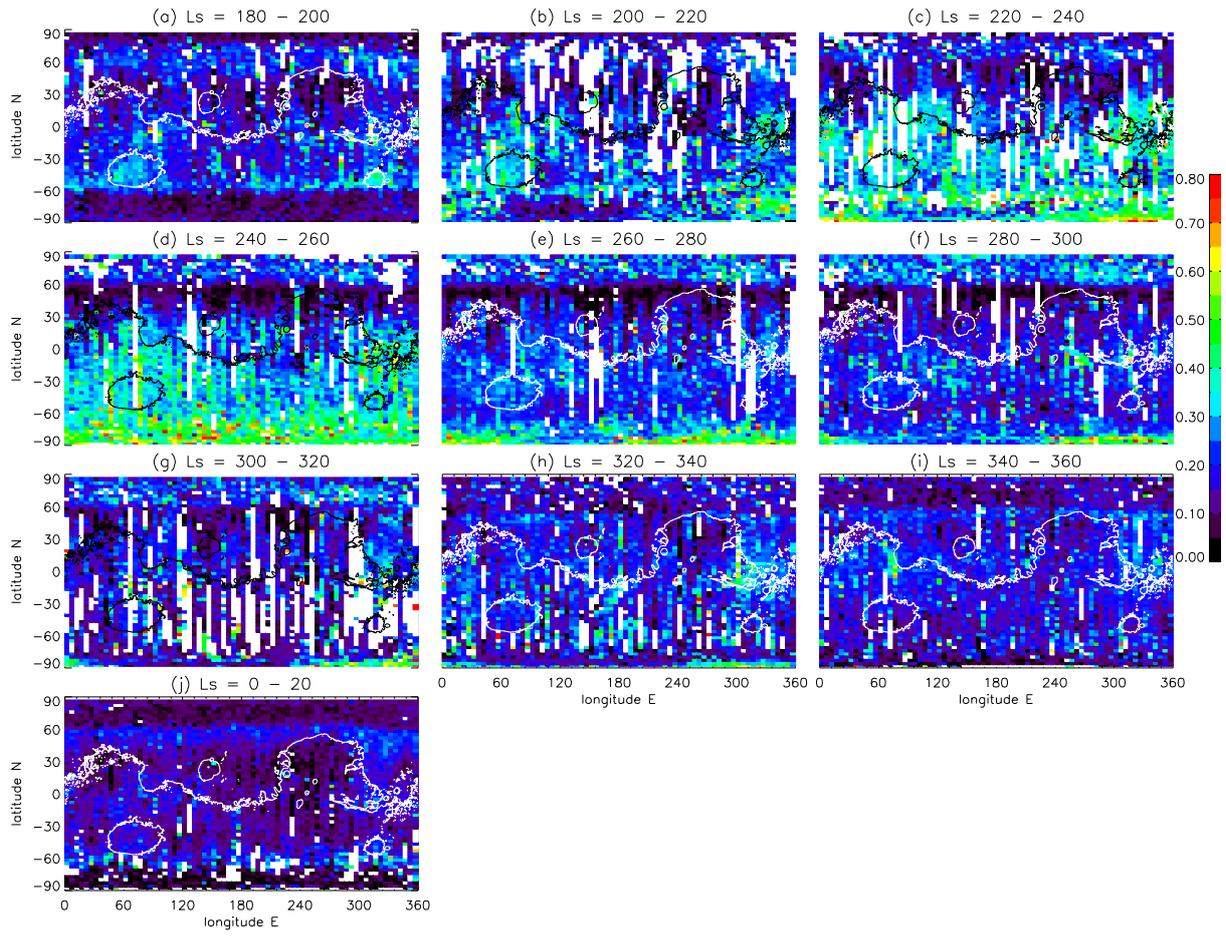

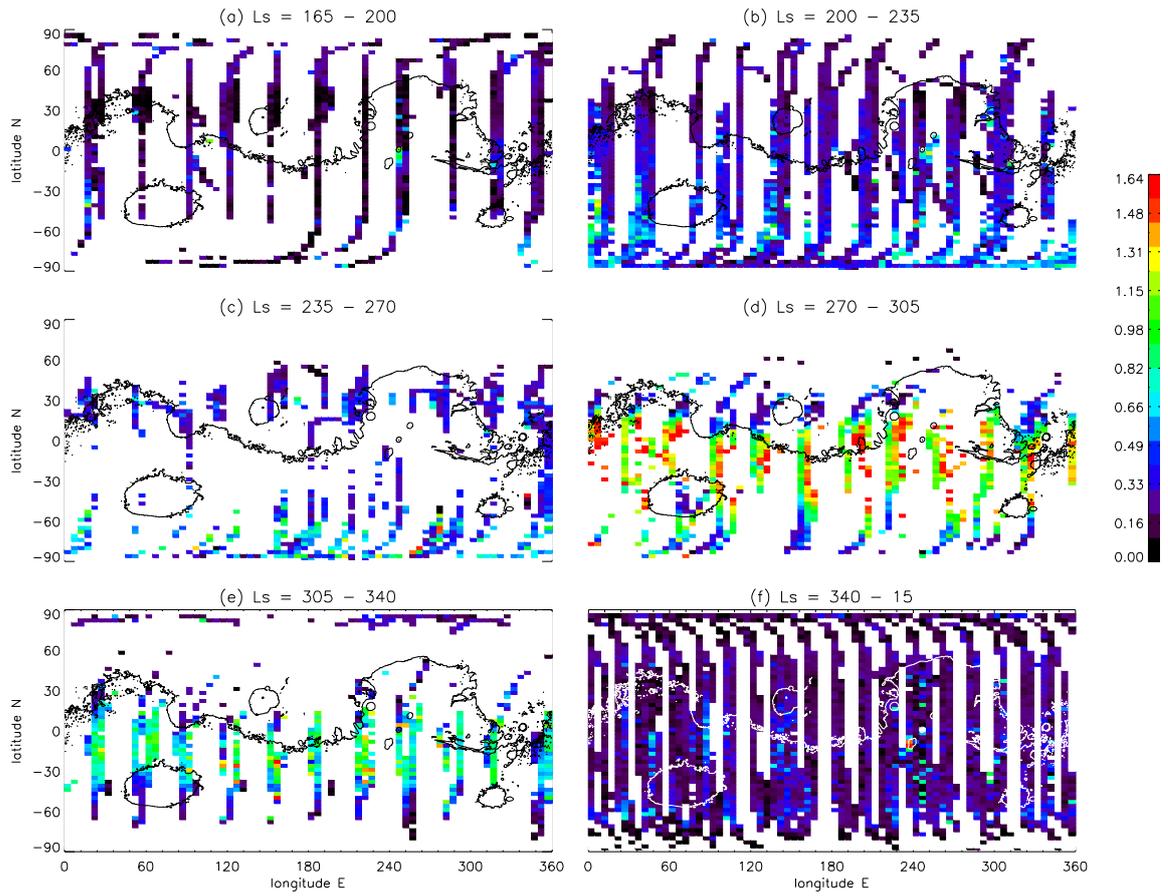

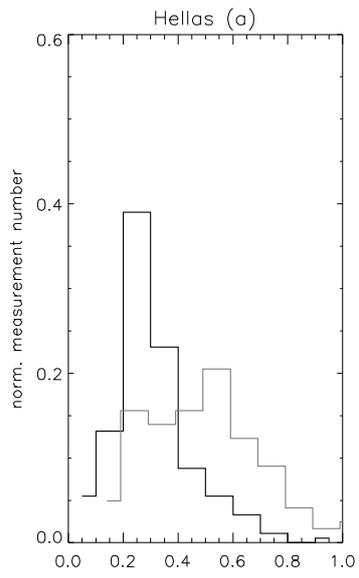 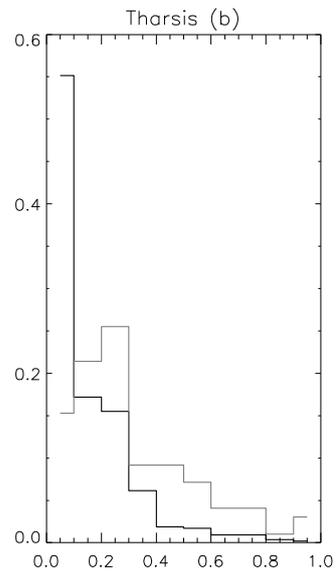 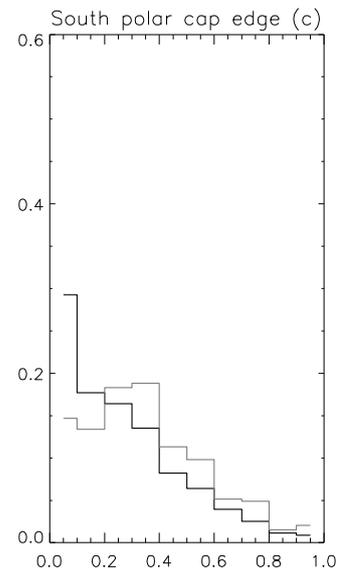

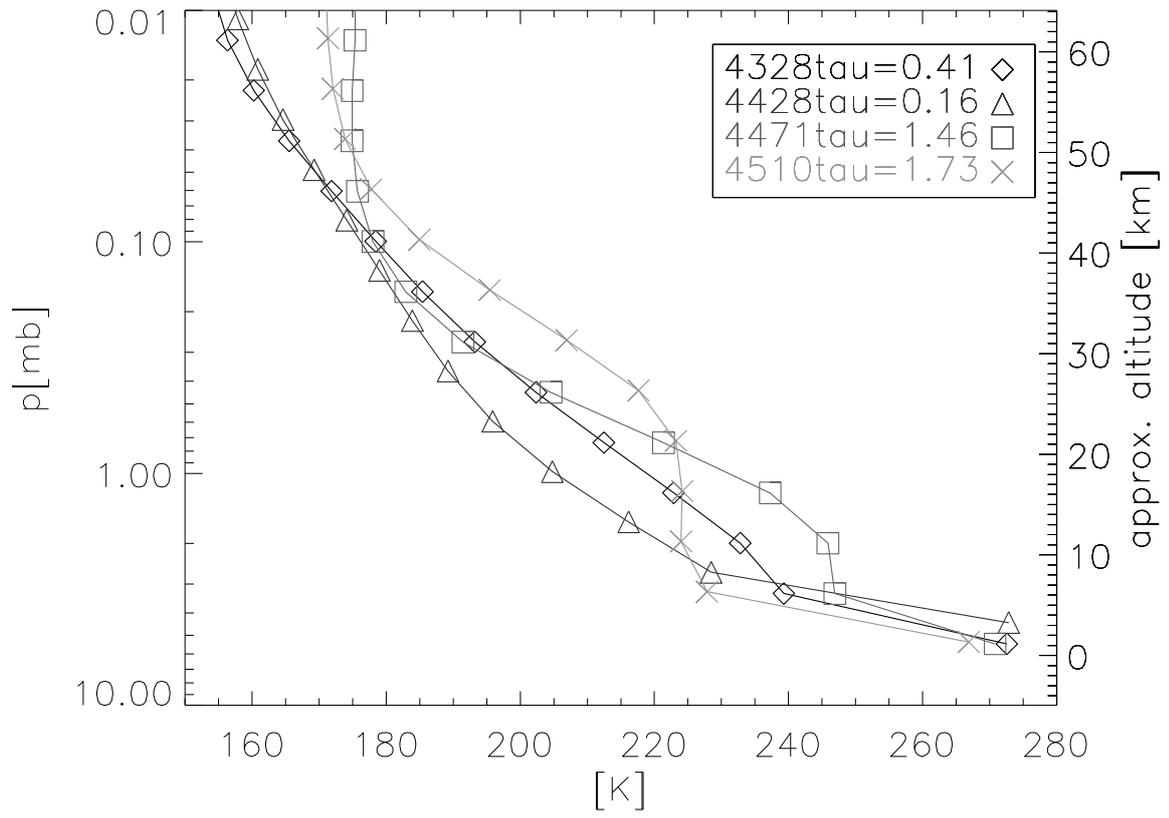

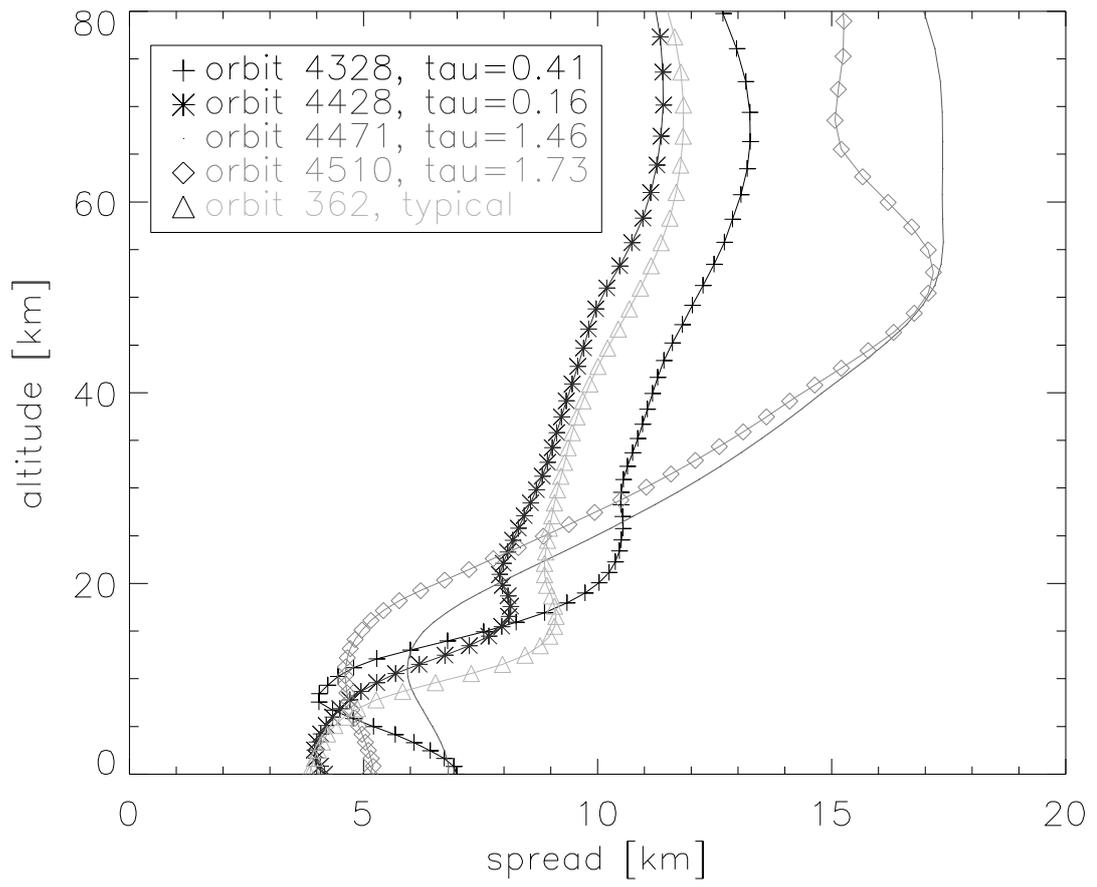

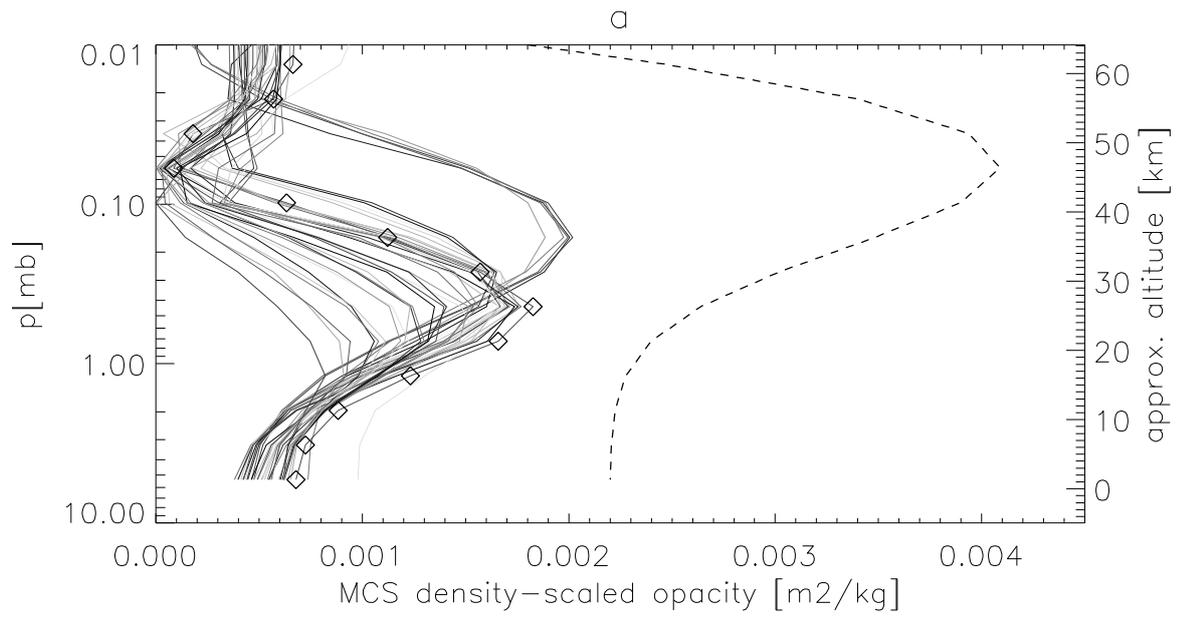

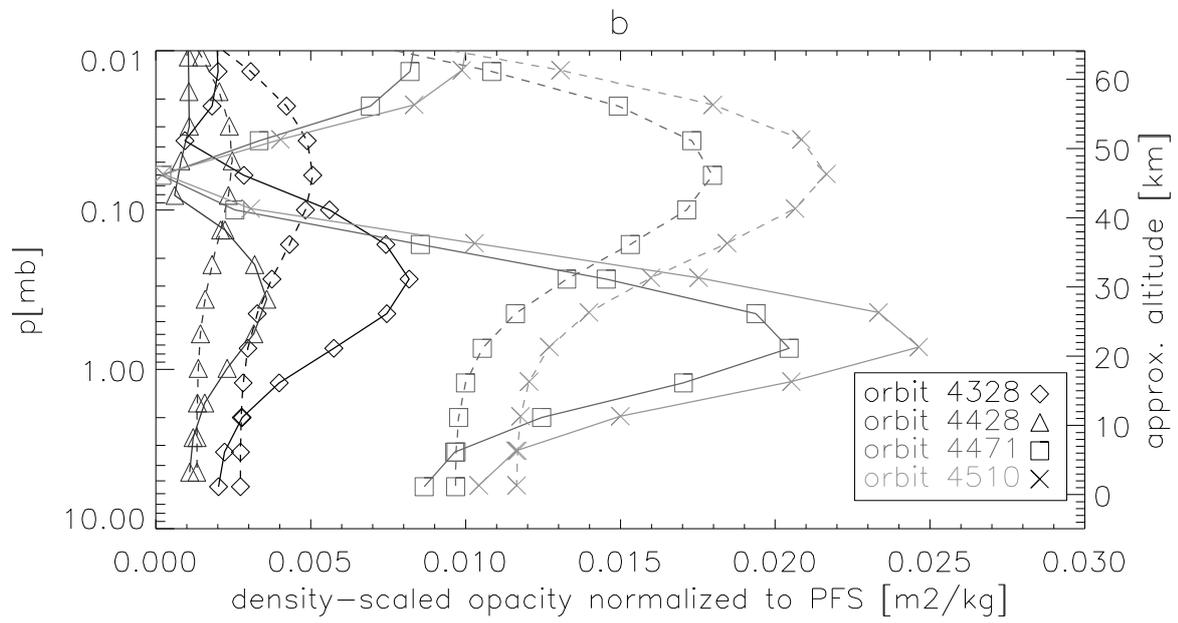

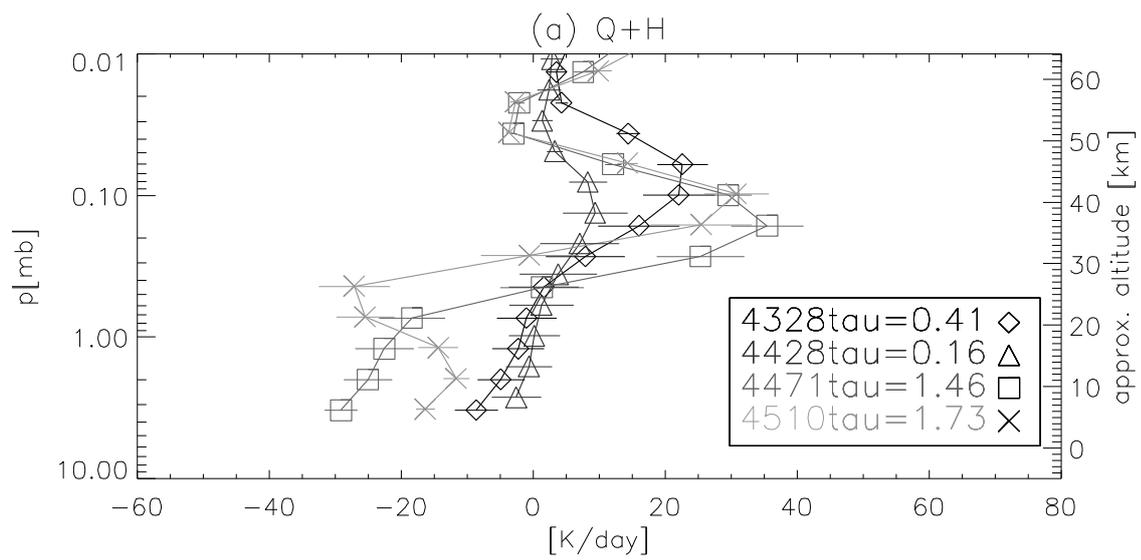

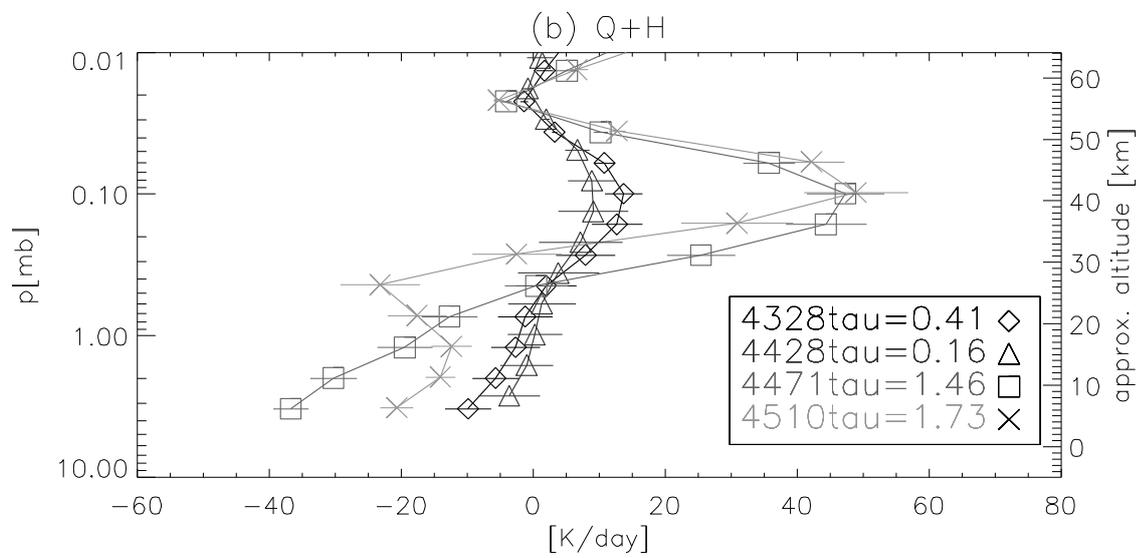

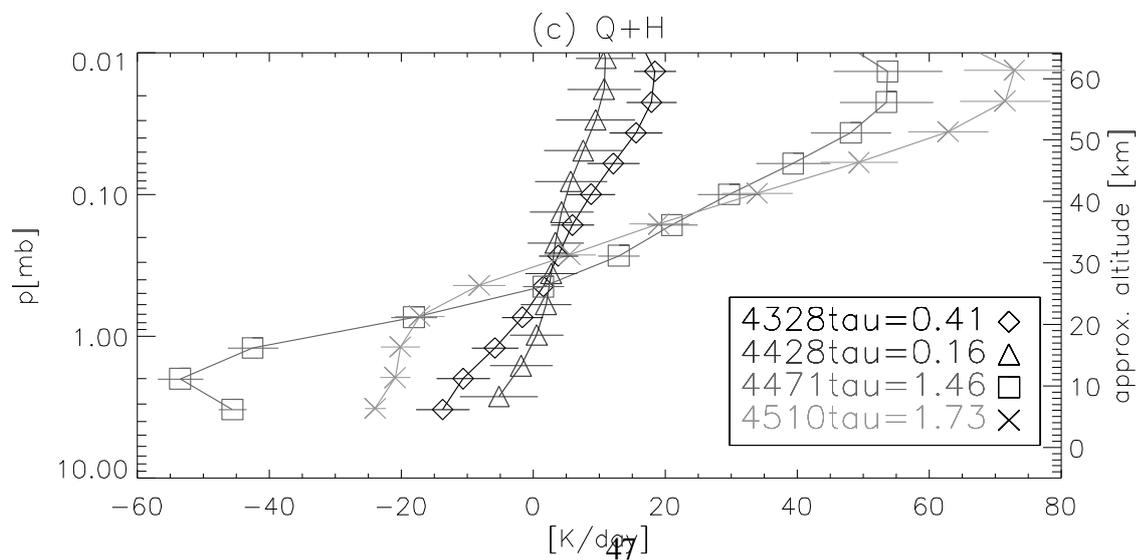